\documentclass[final,1p,times]{elsarticle}
\usepackage{graphicx}
\usepackage{amssymb}   

\journal{Physics Letters A}
\begin{document}

\begin{frontmatter}

\title{Finite Difference-Time Domain solution of Dirac equation and the Klein Paradox}

\author{Neven Simicevic \fnref{1}}
\ead{neven@phys.latech.edu}
\fntext[1]{Correspondence should be addressed to Louisiana Tech University, 
PO Box 10348, Ruston, LA 71272, USA, Tel: +1.318.257.3591, Fax: +1.318.257.2777}
\address{%
Center for Applied Physics Studies, Louisiana Tech University, Ruston, Louisiana 71272, USA\\
}

\date{\today}

\begin{abstract}
The time-dependent Dirac equation is solved using the three-dimensional
Finite Difference-Time Domain (FDTD) method. The dynamics of the electron wave packet 
in a scalar potential is studied in the arrangements associated with the Klein paradox:
potential step barriers and linear potentials. No Klein paradox is observed.

\end{abstract}


\end{frontmatter}


In our previous paper \cite{Sim08}, the Finite Difference-Time Domain (FDTD) method, 
originally introduced by
Kane Yee \cite{Yee66} to solve Maxwell's equations, was for the first time 
applied to solve the three-dimensional Dirac equation. The $Zitterbewegung$ and 
the dynamics of a well-localized electron were used as examples of FDTD 
applied to the case of free electrons. In this paper the Finite Difference-Time Domain 
method is applied to the case of the motion of an electron wave packet 
inside and scattering from the potential step barrier or linearly dependent potential.

In the analysis of the Dirac equation, Oskar Klein calculated that if the electrons scatter from the
repulsive potential step of sufficient strength, they were partially and without impedance transmitted 
even if the transmission was forbidden by the conservation of energy \cite{Klein29}. 
At the same time, it was possible that the reflected current was larger than the incoming one. 
Since the consequences of such a 
finding are counterintuitive and lead to interpretational problems of the Dirac equation,
the effect is known as the Klein paradox. For comparison, if the same 
scattering was analyzed in the non-relativistic limit by solving the Schrodinger equation, the total 
reflection of electrons was obtained, as intuitively expected. 

Despite its counterintuivity, the Klein paradox
profoundly influenced the development of relativistic
quantum mechanics becoming one of its fundamentals. It facilitated the transition 
from the single-particle interpretation of the
Dirac equation into interpreting the Dirac field as a many-body problem \cite{Grein85}. Much
work in nuclear, particle, and astro-physics, not all consistent with each other, has been 
published on the Klein paradox. A recent historical 
review of the paradox was written by Dombey and Calogeracos \cite{Domb99}. Krekora {\it et al.}, 
in addition to publishing the results on their temporally resolved numerical solution 
of the Dirac equation, provided the status of recent theoretical investigation 
\cite{Krek05,Krek05a,Krek04,Krek04a}. While most of the work conforms with the Klein solution and 
the interpretation of the Klein paradox, there is still some dissent, with the most
recent contained in Ref. \cite{Bos07,Drag09,Bow08}. 
Recently, the interest in the Klein paradox increased with the proposal to experimentally test it
in a simple experiment using electrostatic barriers in single- and bi-layer graphene \cite{Kats06},
but until now, no conclusive experimental evidence of the Klein paradox has been found \cite {Drag09}.

The Klein paradox results from the solution of the stationary time-independent Dirac equation. 
In this work, the results of the solution of the time-dependent Dirac equation are presented. 
The FDTD scheme is applied to the Dirac equation for the case when the
electromagnetic field described by the four-potential ${A^{\mu}=\{A_{0}(x),\vec A(x)\}}$
is minimally coupled to the particle \cite{Grein85,Sak87}
\begin{equation}
{\imath \hbar {\frac{\partial \Psi}{\partial t}}= ({H}_{free}+{H}_{int}) \Psi},
\label{Dirac_eq}
\end{equation}
where
\begin{equation}
{{H}_{free} = -\imath c\hbar {{\bf\alpha} \cdot \nabla} + \beta m c^{2}},
\end{equation}
\begin{equation}
{{H}_{int} = - e {{\bf\alpha} \cdot {\vec A} } + e A_{0}},
\end{equation}
and
\begin{equation}
{\Psi (x) =\left( \begin{array} {c} \Psi_{1} (x) \\ \Psi_{2} (x)
\\ \Psi_{3} (x)\\ \Psi_{4} (x) \end{array} \right)}.
\end{equation}
The matrices ${\bf\alpha}$ and $\beta$ are expressed using $2 \times 2$ Pauli
matrices $\bf\sigma^{'}s$ and the $2 \times 2$ unit matrix $I$.

The FDTD schematics to solve
Eq. (\ref{Dirac_eq}) follows Yee's leapfrog algorithm \cite{Sim08,Yee66}.
The wave functions $\Psi_{1}$ and $\Psi_{2}$ at the time $n-1/2$ are used to calculate the
wave functions $\Psi_{3}$ and $\Psi_{4}$ at the time $n$, which are then used to calculate
the wave functions $\Psi_{1}$ and $\Psi_{2}$ at the time $n+1/2$, and so on. 
The same numerical requirements as in the case of electrodynamics are followed \cite{Sim08}. 
While the dynamics of a Dirac electron can be studied
in an environment described by any four-potential $A^{\mu}$ regardless
of its complexity and time dependency, the study of the Klein paradox requires only that
$A_{0}(x) \neq 0$ in the designated region and $\vec A(x) = 0$ everywhere.
 
As a consequence of the Dirac equation being of first order and linear
in $\partial / \partial t$, the entire dynamics of the electron is defined,
as in the case of Maxwell's equations, only by its
initial wave function. The dynamics
of a wave packet is defined by its initial wave function
\begin{equation}
{\Psi (\vec x,0) =N \sqrt{\frac{E+mc^{2}}{2E}}\left( \begin{array} {c} 1 \\ 0
\\ \frac{p_{3}c}{E+mc^{2}}\\ \frac{(p_{1}+ip_{2})c}{E+mc^{2}}\end{array} \right)}
e^{-\frac{\vec x \cdot \vec x }{4x_{0}^{2}}+\frac{i\vec p \cdot \vec x}{\hbar}},
\label{Wave_packet}
\end{equation}
where $N=[(2\pi)^{3/2}x_{0}^{3}]^{-1/2}$ is normalizing constant. Eq. (\ref{Wave_packet})
represents a wave packet whose initial probability distribution is of a normalized Gaussian
shape. Its size is defined by the constant $x_{0}$, its spin is pointed along the z-axis, and its
motion is defined by the values of $p_{1},p_{2}$, and $p_{3}$. In the
``single particle interpretation" of the Dirac equation, if we choose $p_{2}=p_{3}=0$ and 
$p_{1} \neq 0$ the wave packet  should move in the x-direction. This is not the case. 
Because of the localization of the wave packet
and limitation of the direction of the spin, the initial condition in
Eq. (\ref{Wave_packet}) may contain also 
a component moving in the opposite direction \cite {Huang52,Thal04}.
For the case of free electrons some of the dynamics of this wave packet were studied 
in Ref. \cite{Sim08}.

The dynamics of the wave packet is very complex. The energy of the particle described by the packet 
depends on its localization, defined by the Gaussian component of the wave function, and its
initial momentum, part of the wave function's phase. While an extensive study was done in order to 
understand the interplay between the height and the shape of the potential barrier, and
particle localization and its initial momentum, in this paper, we report only the results of several
cases associated with the Klein paradox: the penetration into a potential barrier of 
supercritical potential satisfying the condition $eV>E+mc^{2}$.
\begin{figure}
\centering
{\scalebox{.35}{\includegraphics{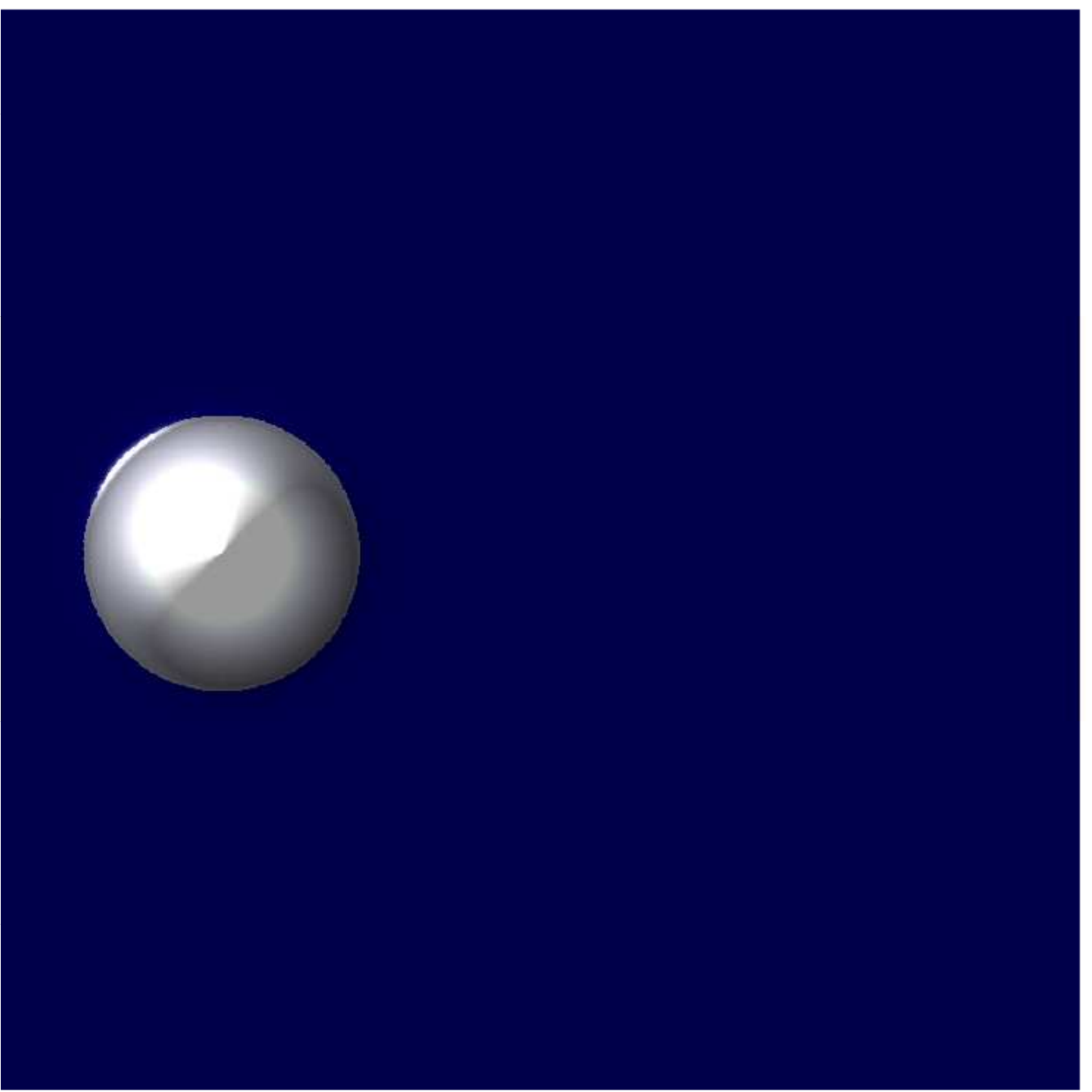}}}
{\scalebox{.35}{\includegraphics{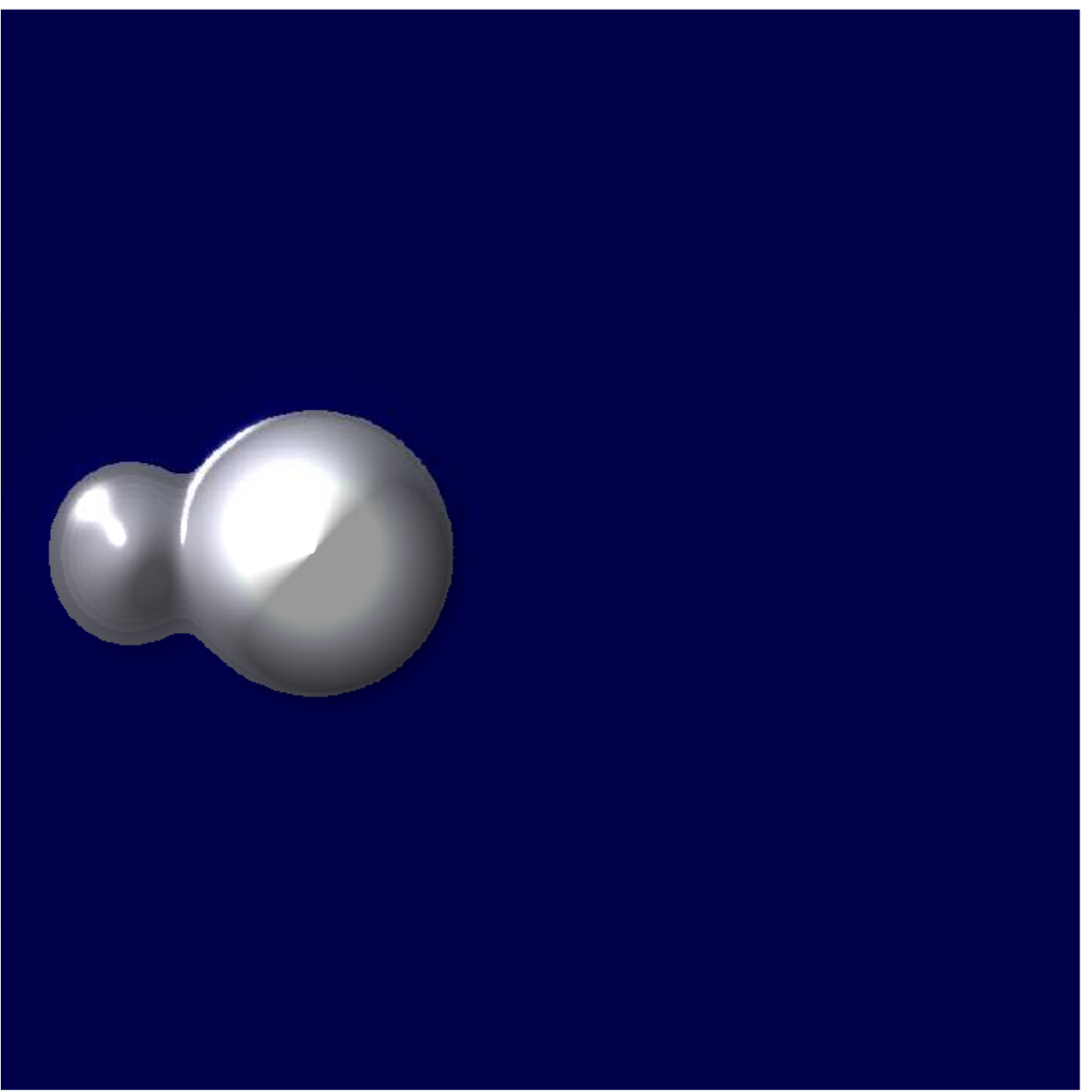}}}
{\scalebox{.35}{\includegraphics{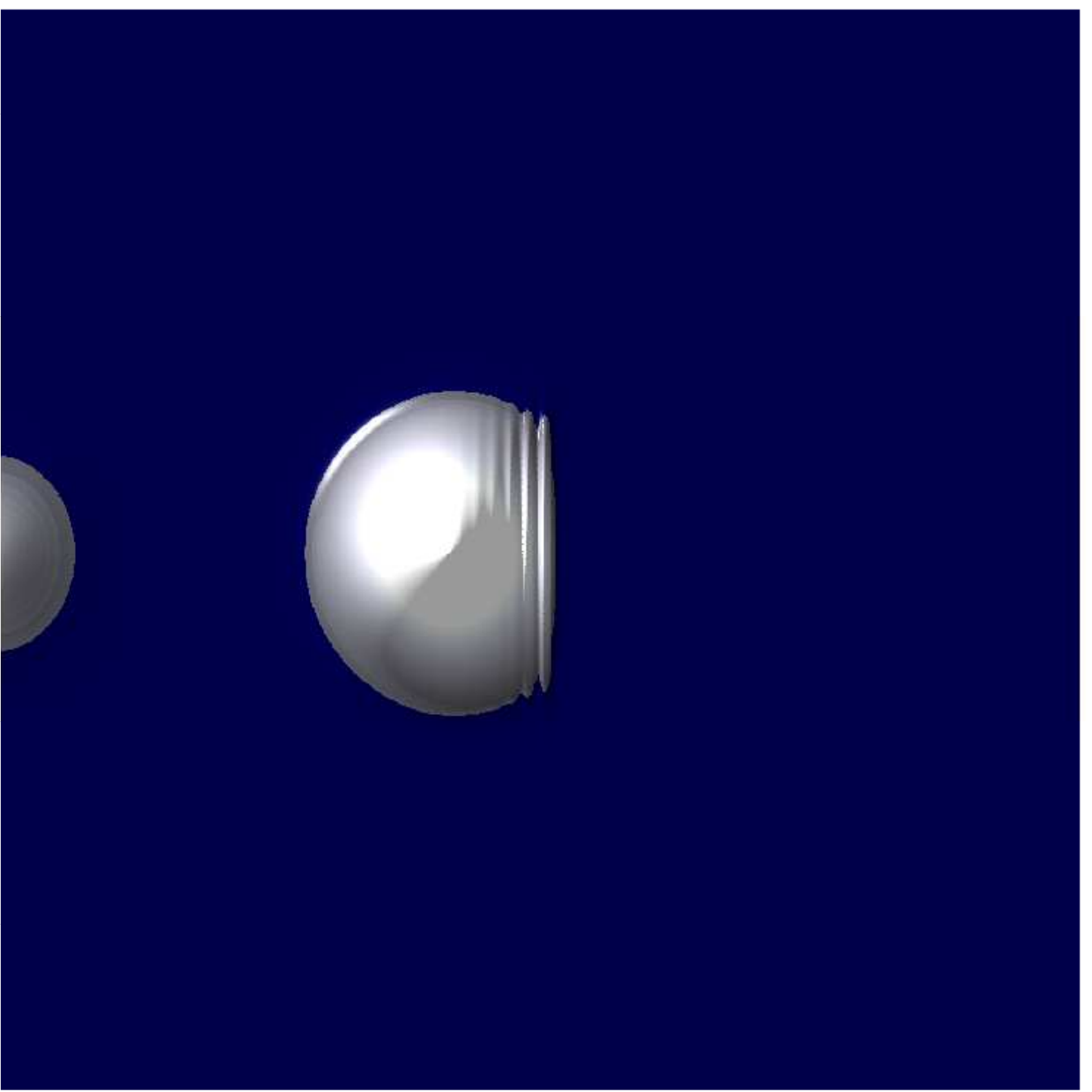}}}
{\scalebox{.35}{\includegraphics{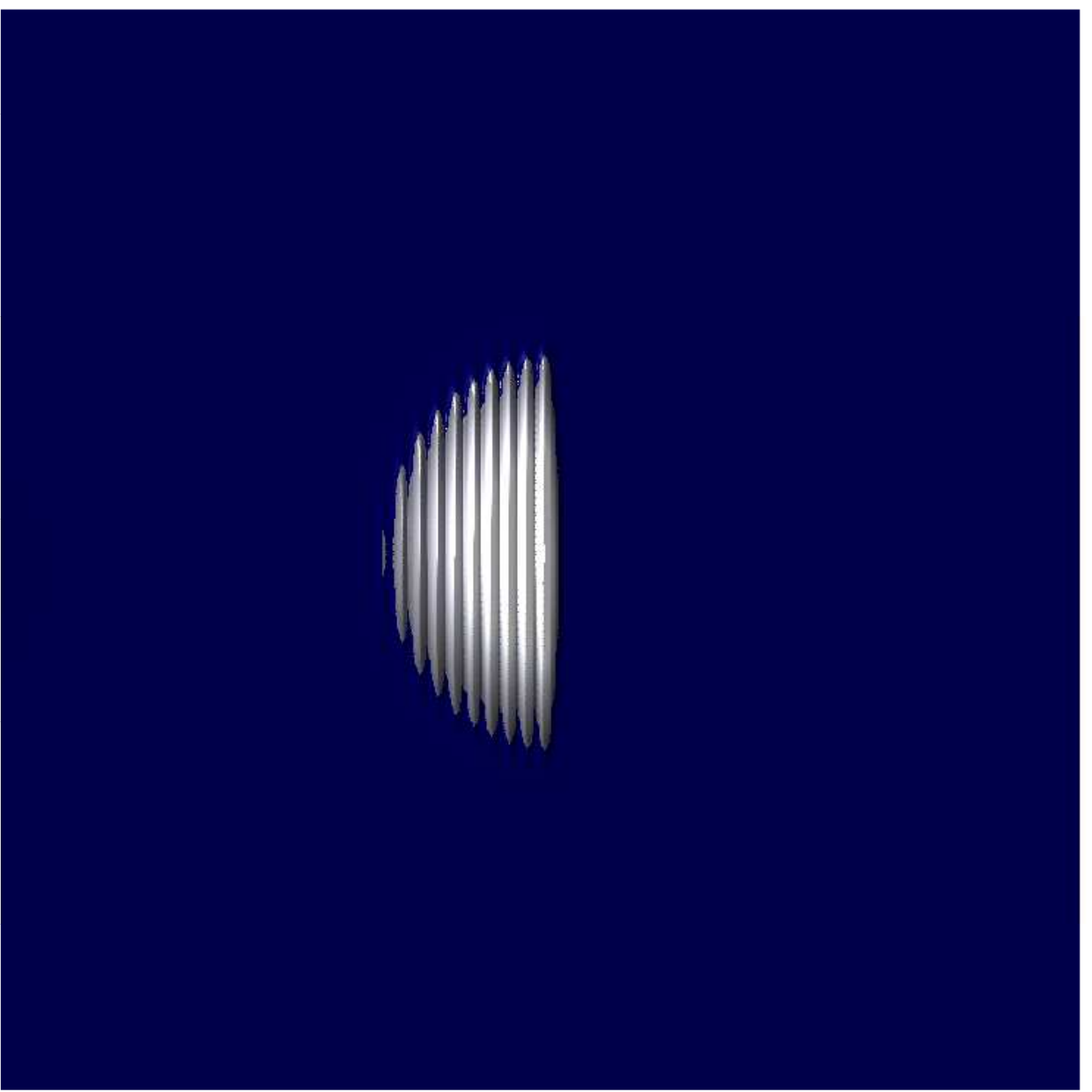}}}
{\scalebox{.35}{\includegraphics{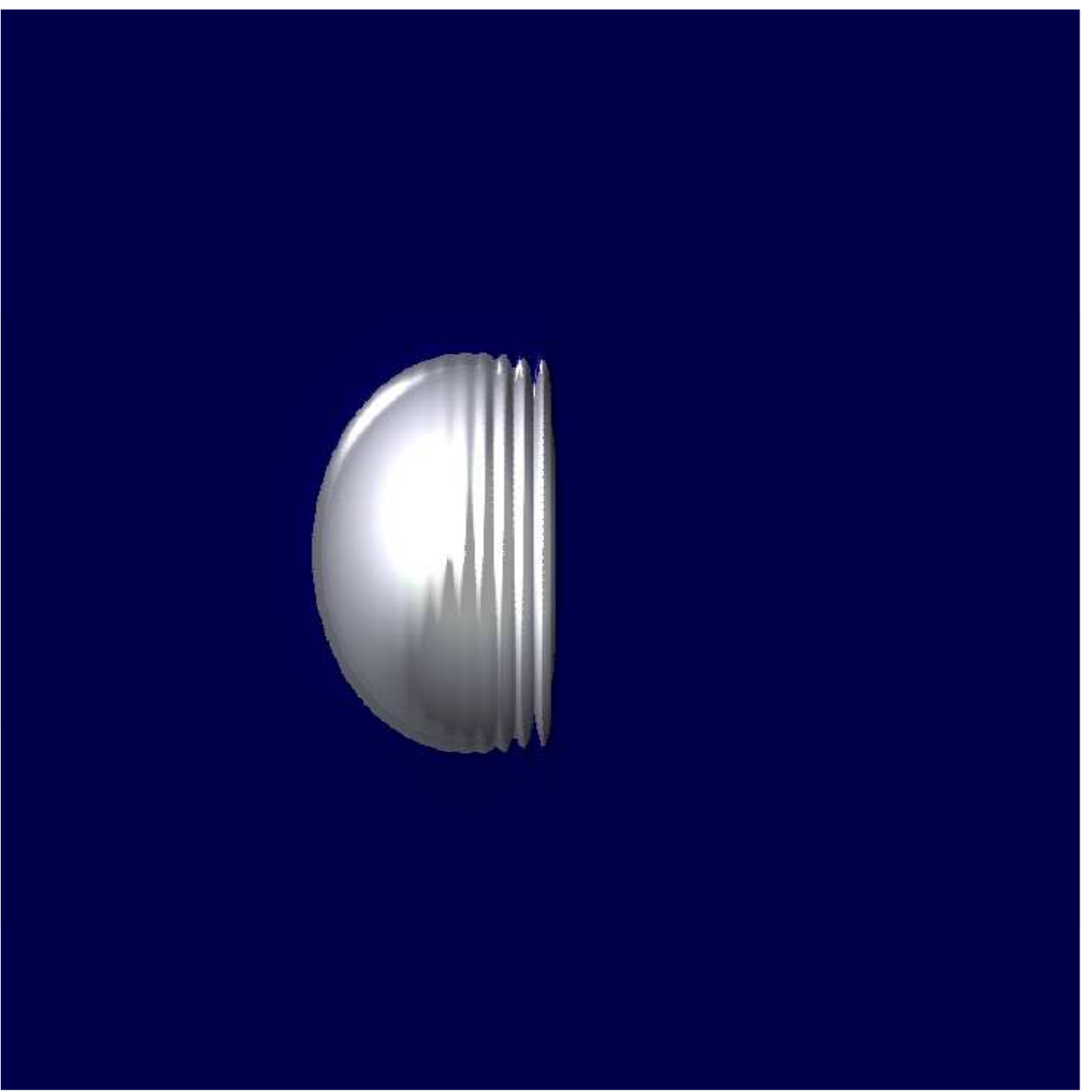}}}
{\scalebox{.35}{\includegraphics{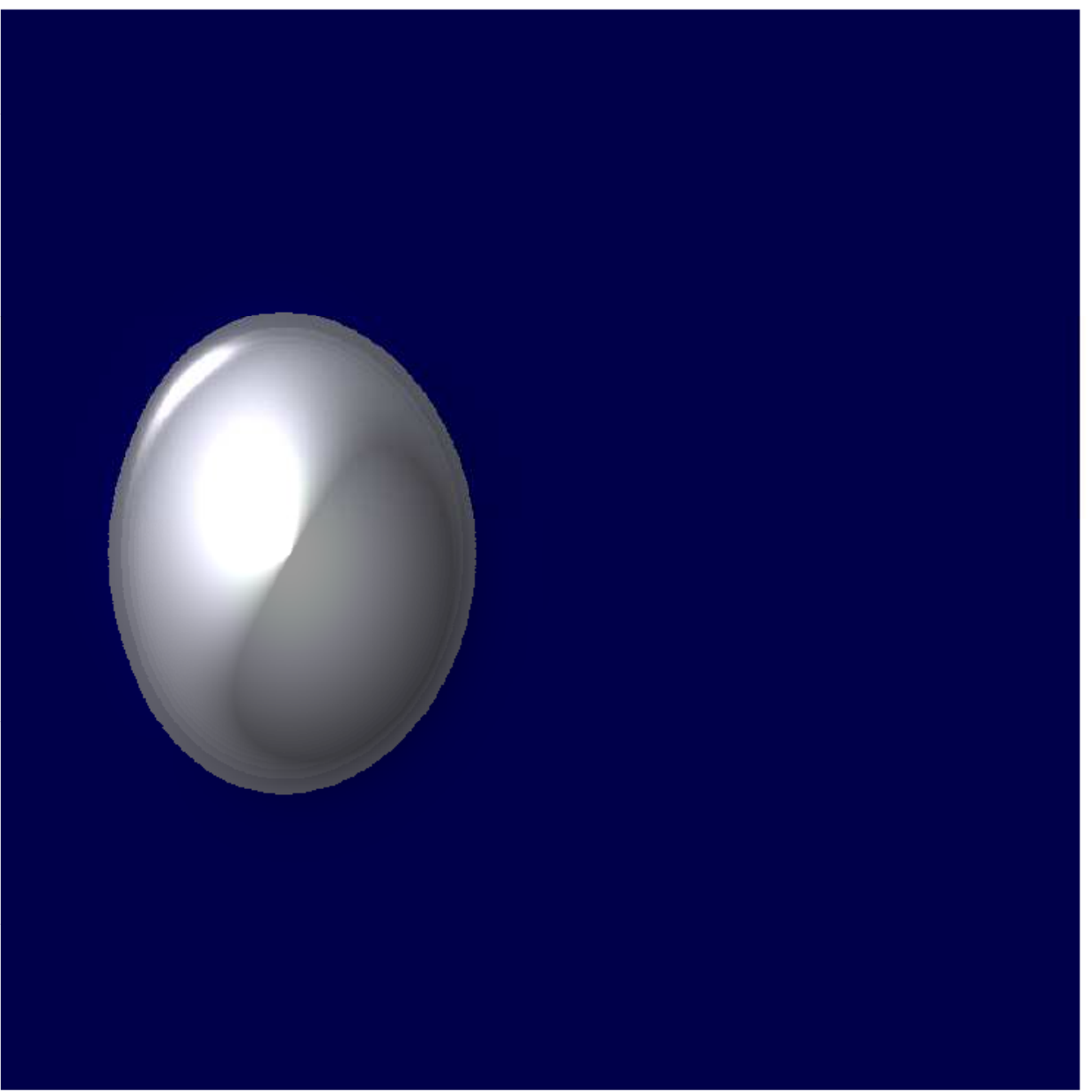}}}
\caption{\label{fig:Wave_repulsive} Six stages of motion of
the wave packet initialized by Eq. (\ref{Wave_packet}) in the x-z plane.
After splitting in two as a result of the initial conditions, the larger component of the wave 
packet moves in the +x-direction and completely reflects from the repulsive supercritical 
potential step barrier. One can observe the interference between incoming and reflecting wave, 
but no Klein paradox. The animation can be accessed on-line \cite{Simi08}.}
\end{figure}

\begin{figure}
\centering
{\scalebox{.35}{\includegraphics{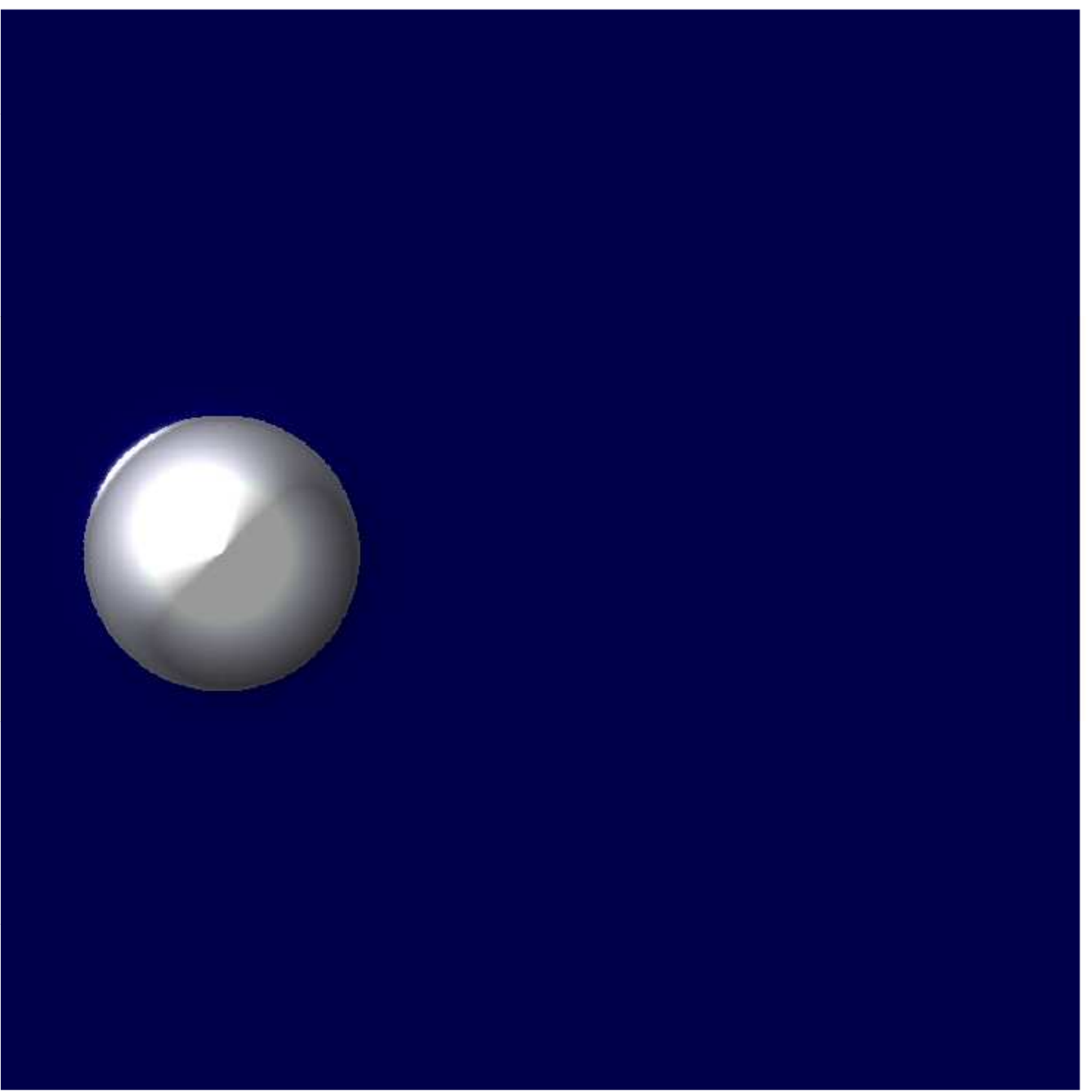}}}
{\scalebox{.35}{\includegraphics{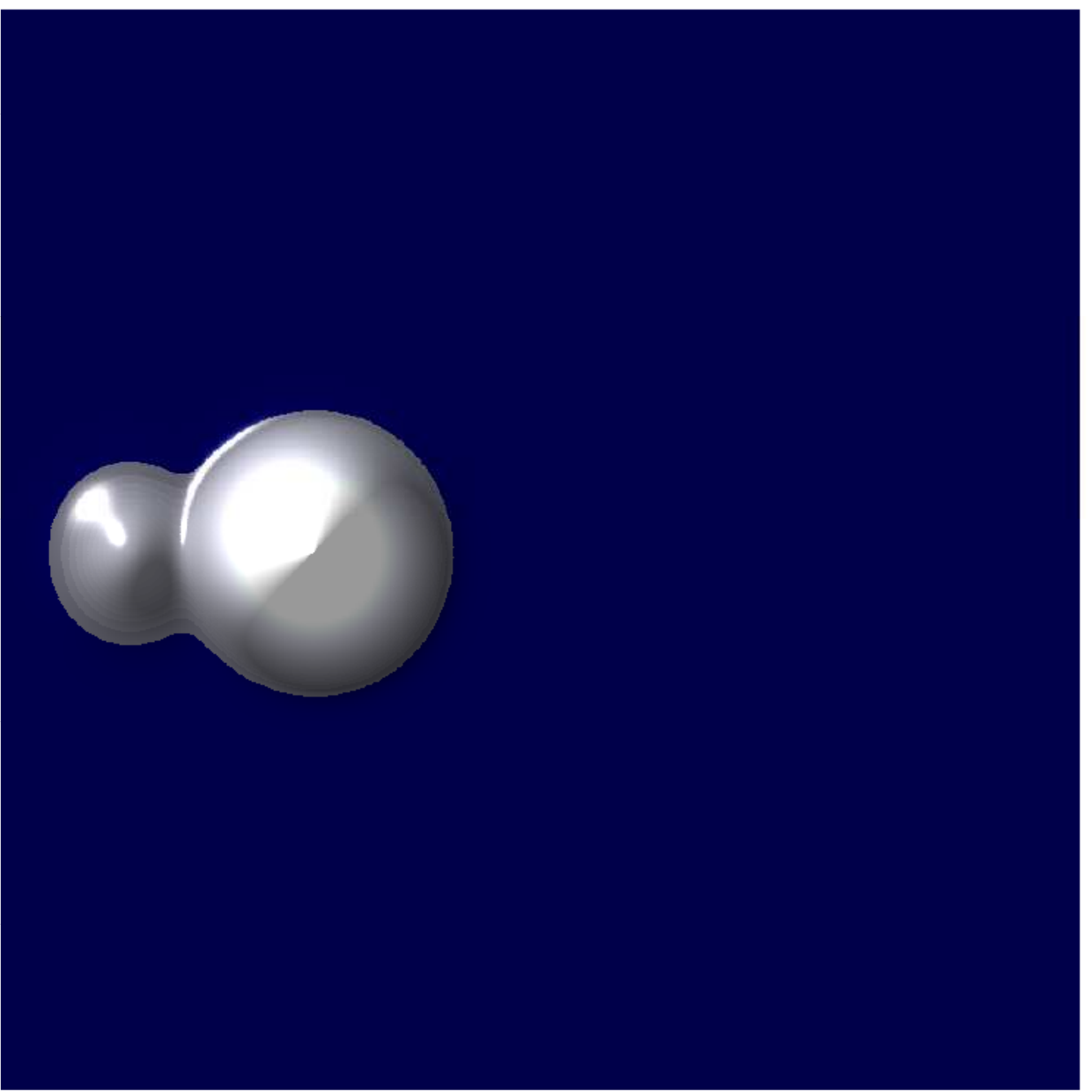}}}
{\scalebox{.35}{\includegraphics{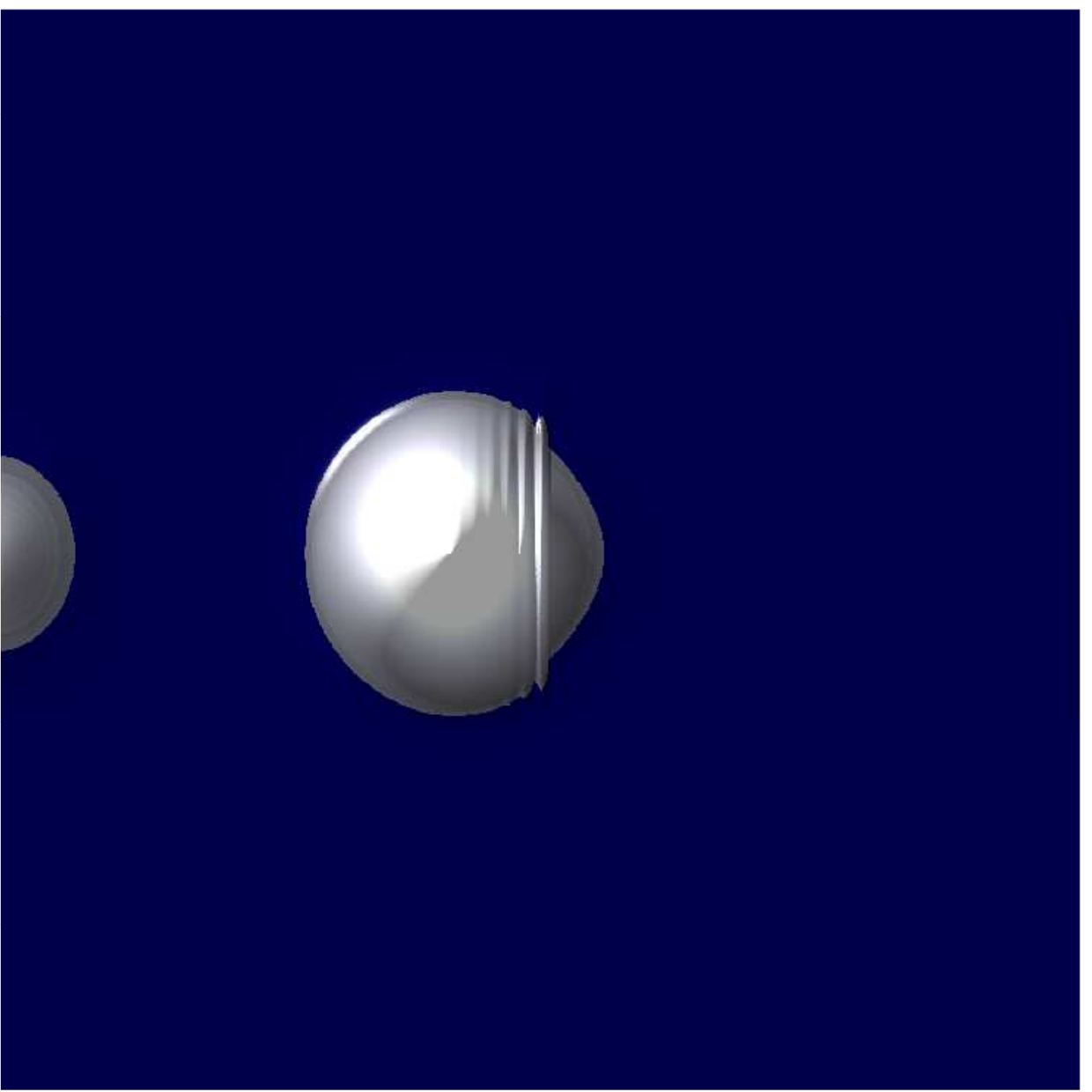}}}
{\scalebox{.35}{\includegraphics{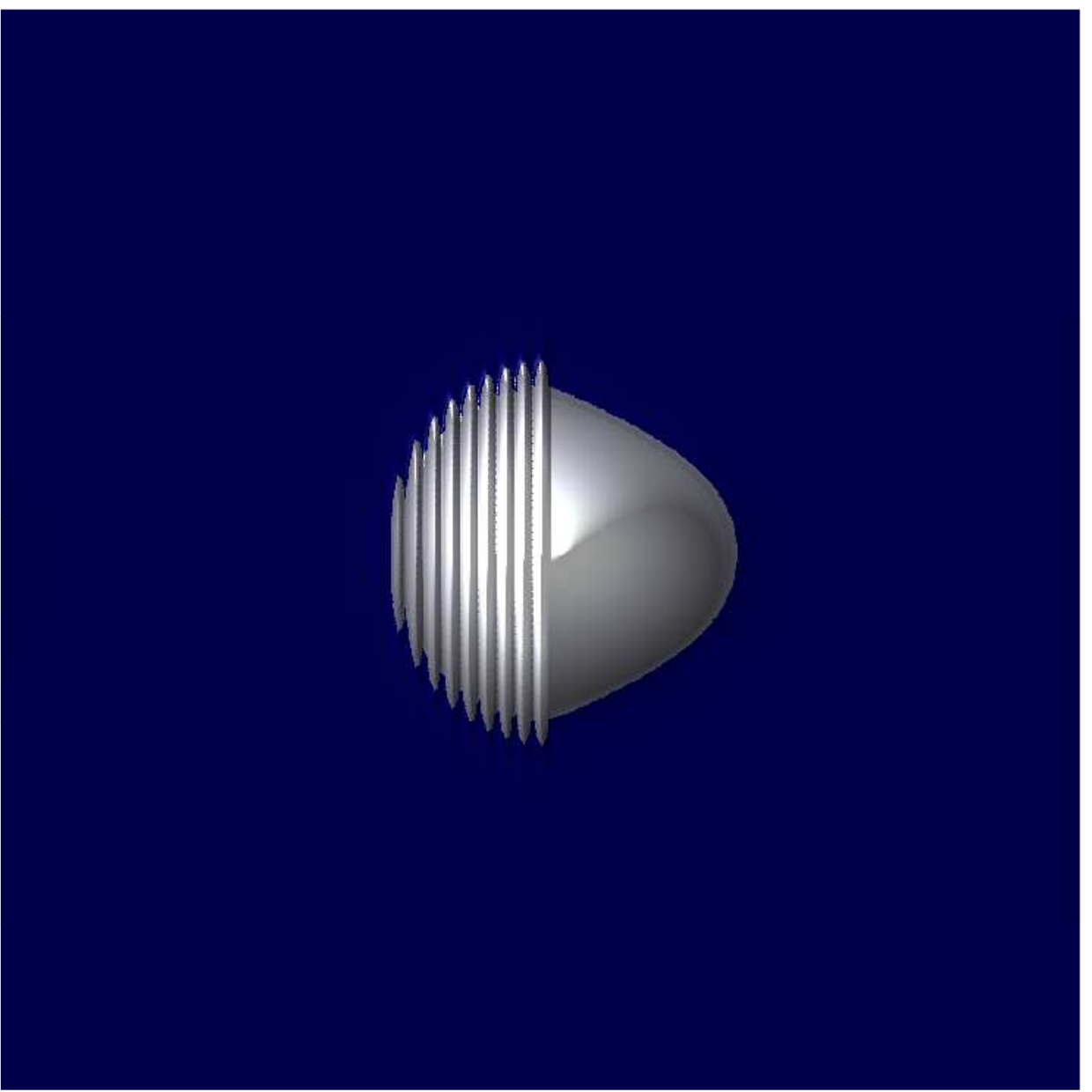}}}
{\scalebox{.35}{\includegraphics{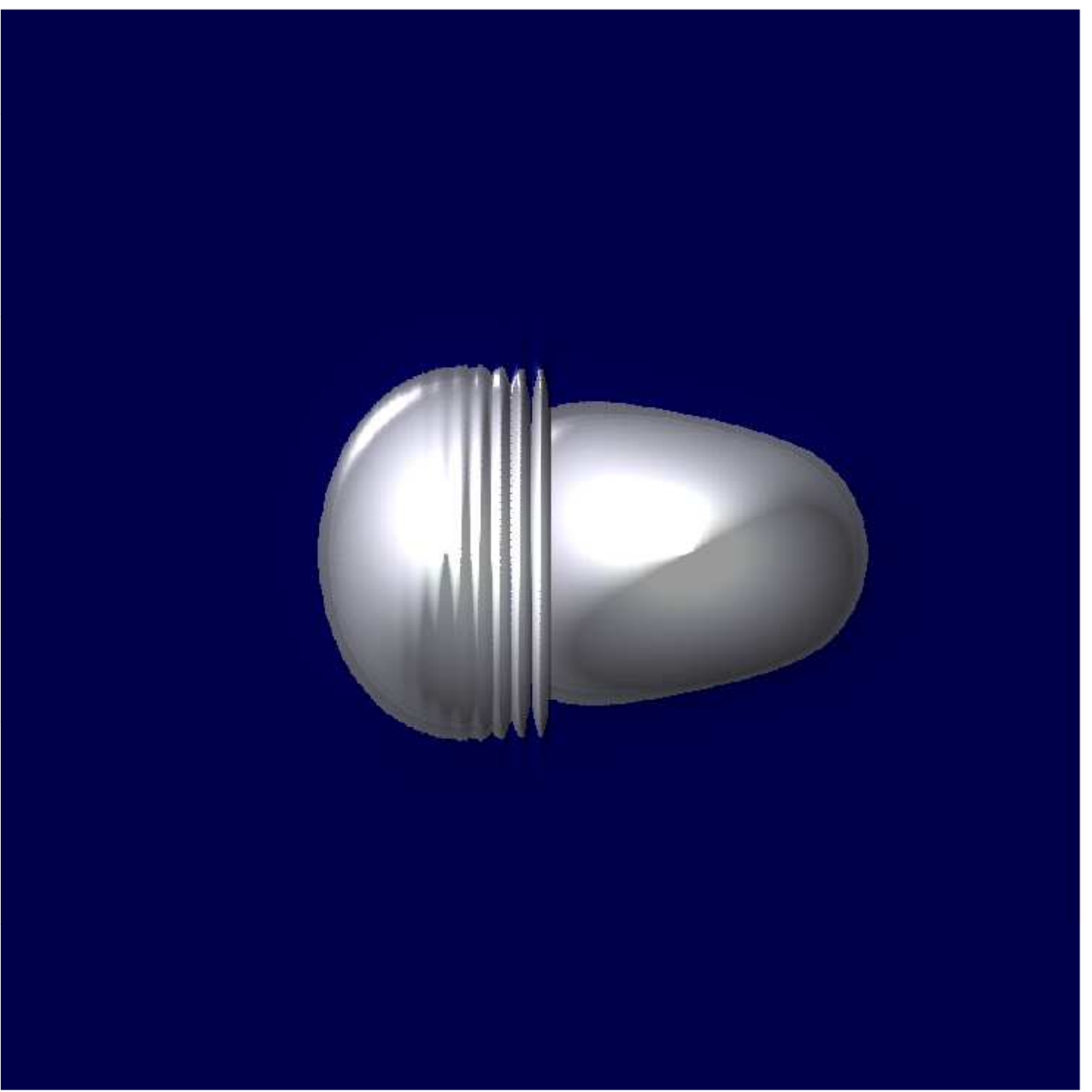}}}
{\scalebox{.35}{\includegraphics{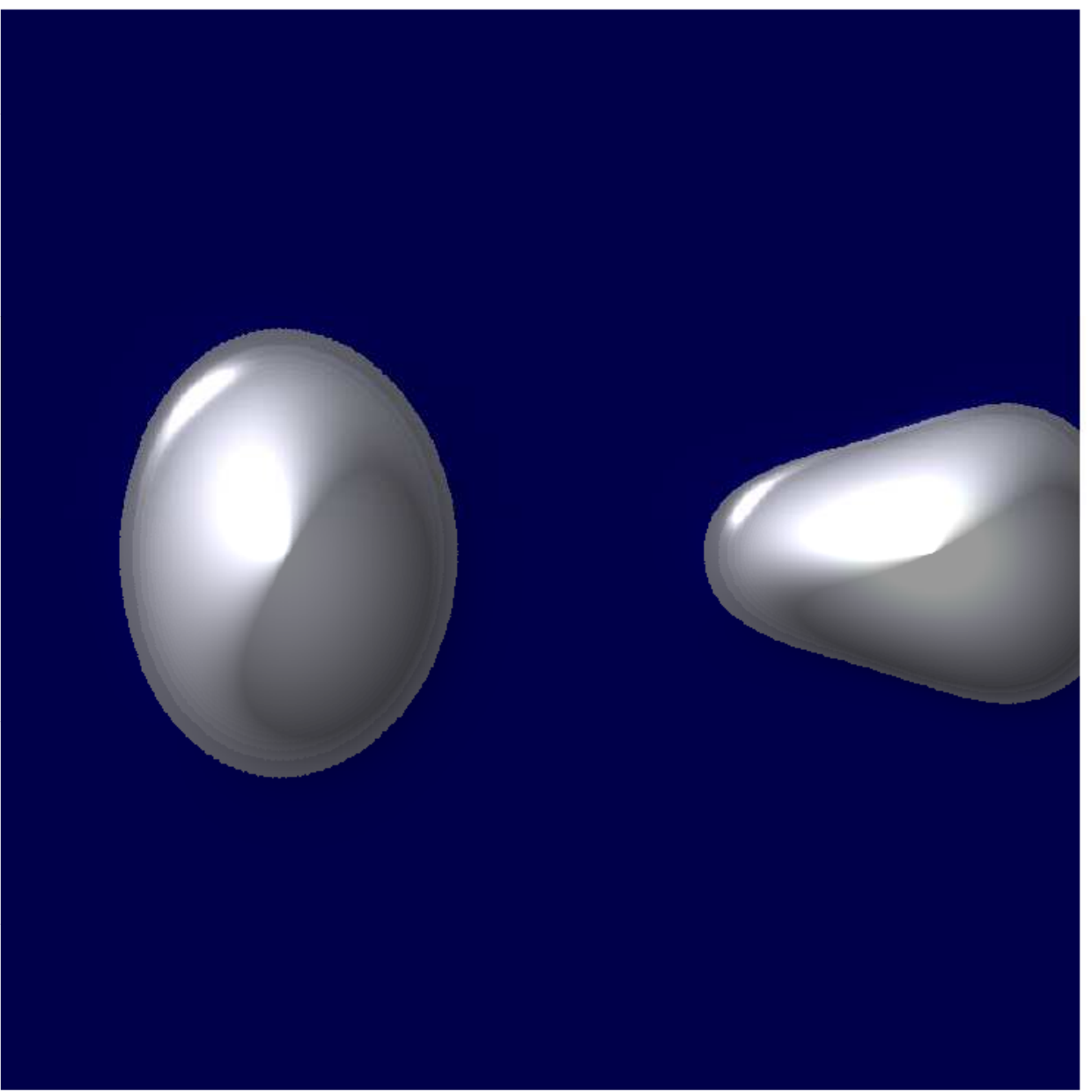}}}
\caption{\label{fig:Wave_attractive} Six stages of motion of
the wave packet initialized by Eq. (\ref{Wave_packet}) in the x-z plane.
The wave packet partially reflects and partially 
penetrates the attractive supercritical 
potential step barrier. In addition to the interference between incoming and reflecting wave, 
the shape of the wave packet changes as it penetrates the barrier. 
The animation can be accessed on-line \cite{Simi08}.}
\end{figure}

\begin{figure}
\centering
{\scalebox{.335}{\includegraphics{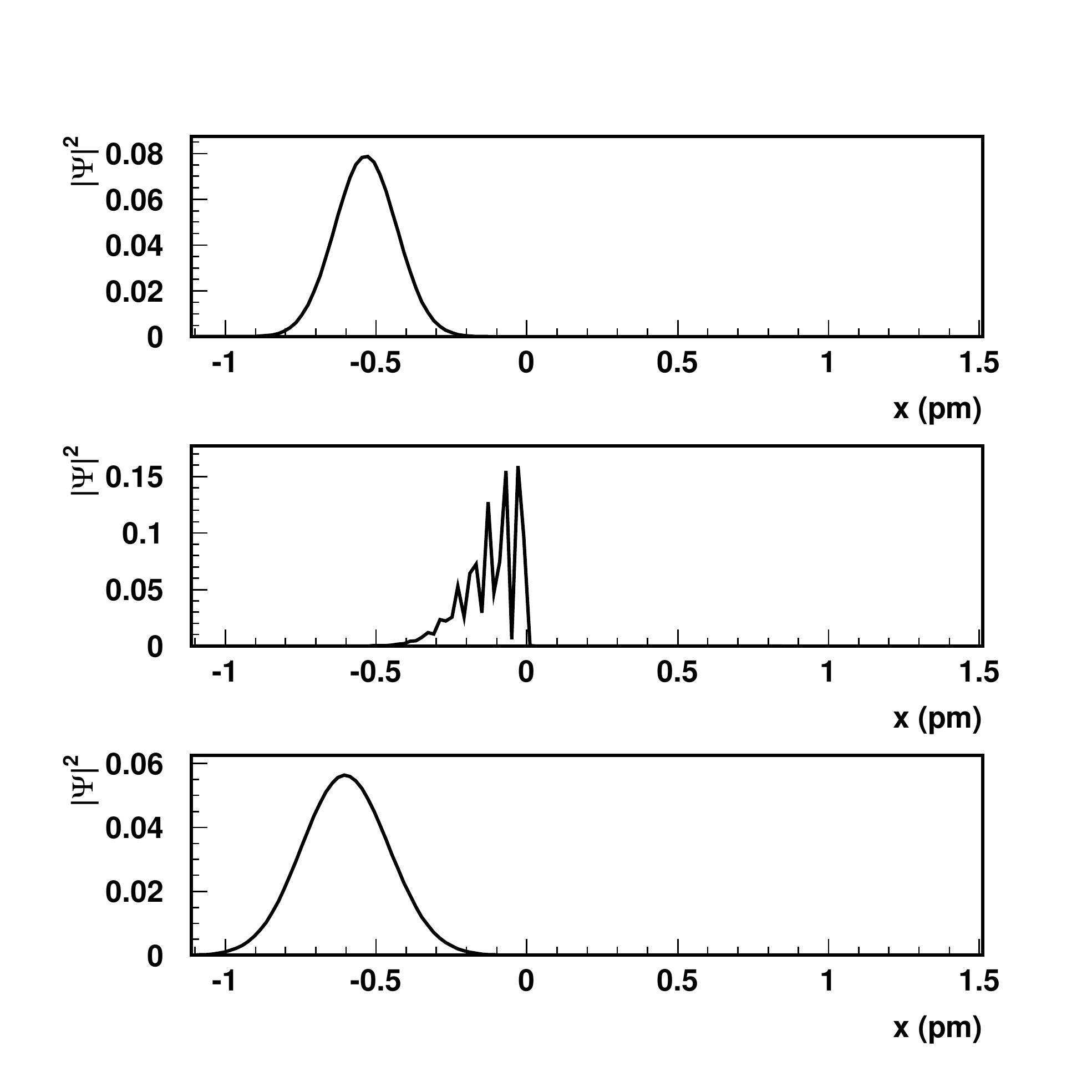}}}
{\scalebox{.335}{\includegraphics{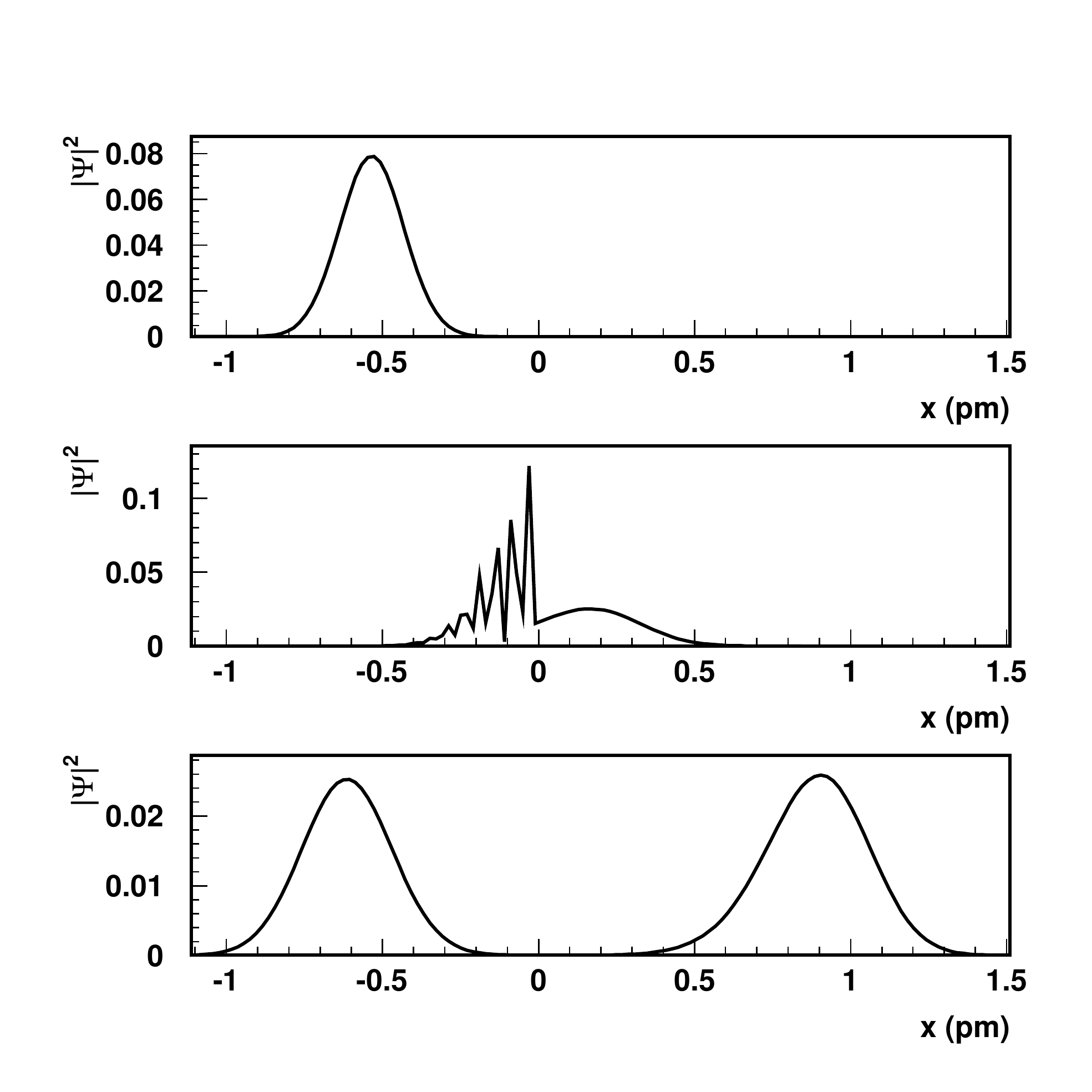}}}
\caption{\label{fig:Psi_rep_att} The probability density function $|\Psi|^2$ for the dynamics 
associated with the repulsive supercritical potential step barrier (left)
and with the attractive supercritical potential step barrier (right). $|\Psi|^2$ is shown
at the time before the wave packet interacts with potential barrier (top), at the 
time of the interaction (in the middle), and at the time when the wave packet is completely reflected 
from the repulsive barrier or is completely separated for the attractive barrier (bottom).} 
\end{figure}

\begin{figure}
\centering
{\scalebox{.35}{\includegraphics{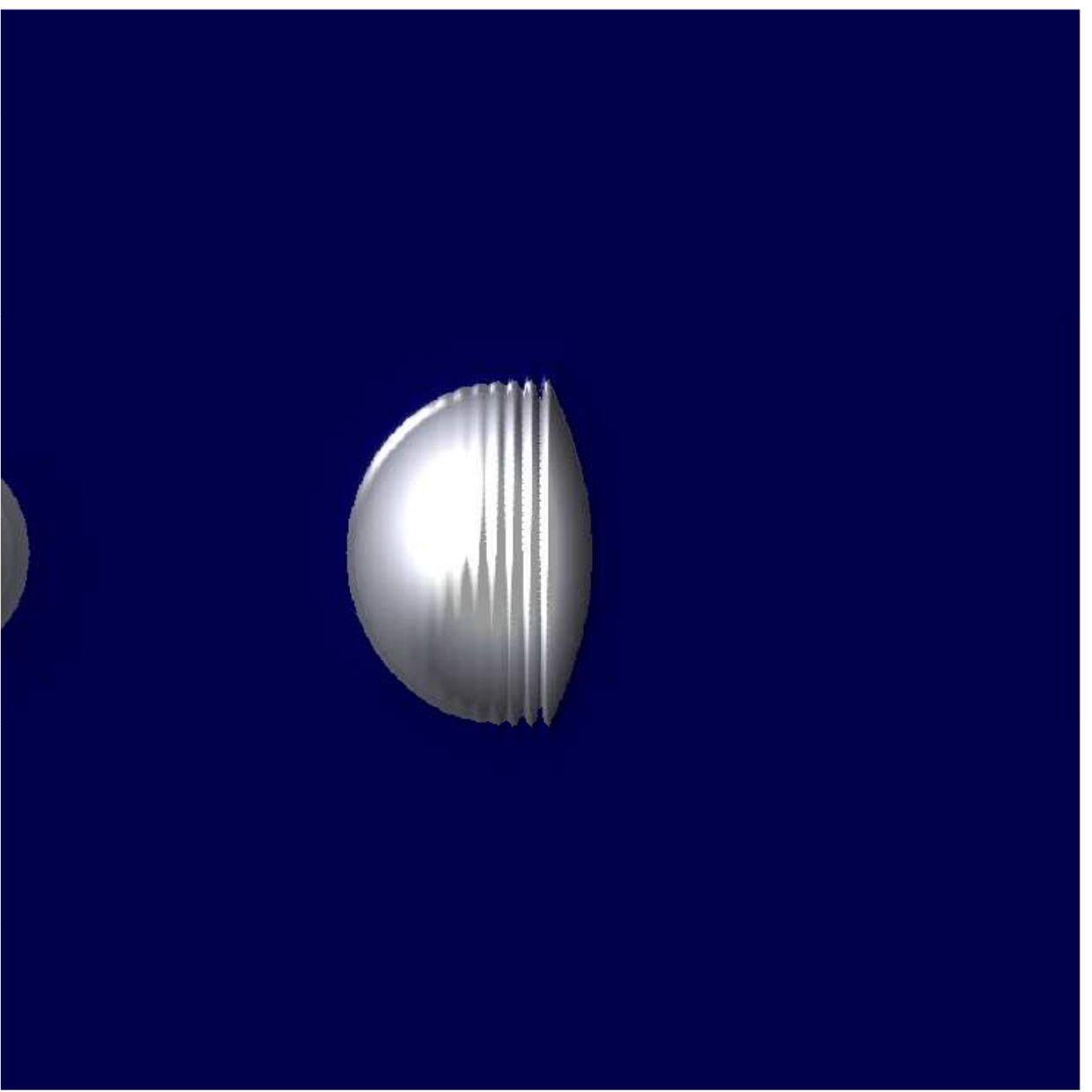}}}
{\scalebox{.35}{\includegraphics{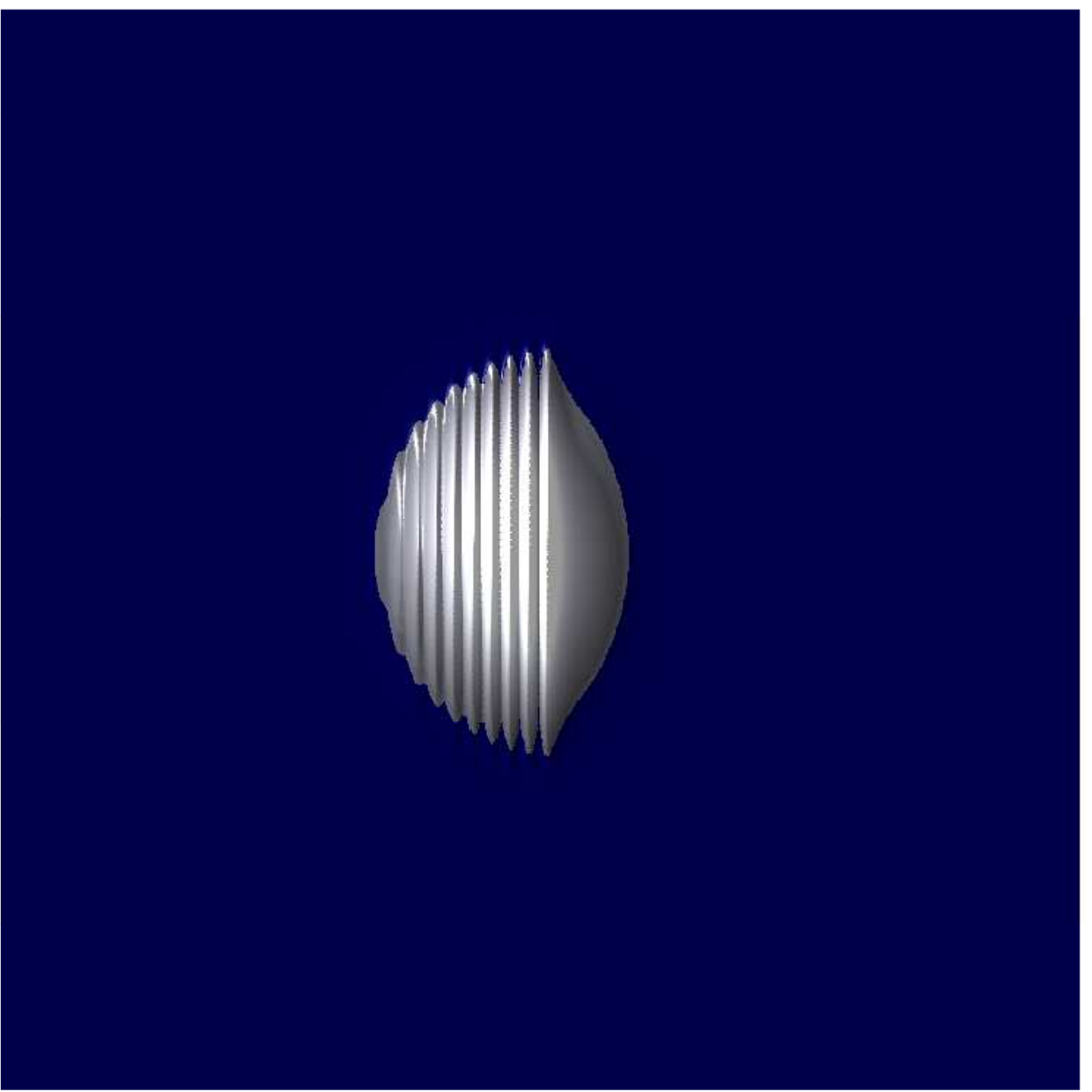}}}
{\scalebox{.35}{\includegraphics{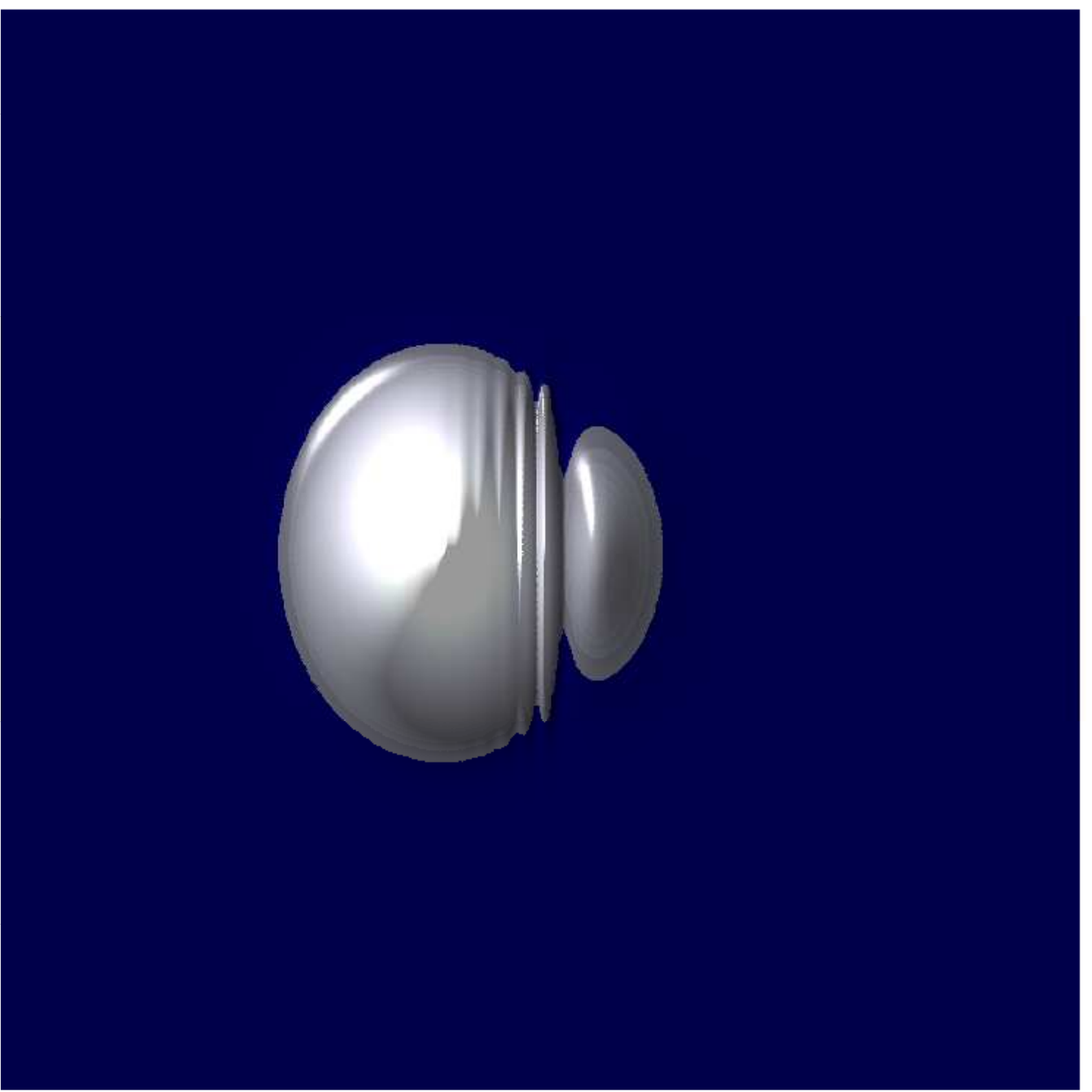}}}
{\scalebox{.35}{\includegraphics{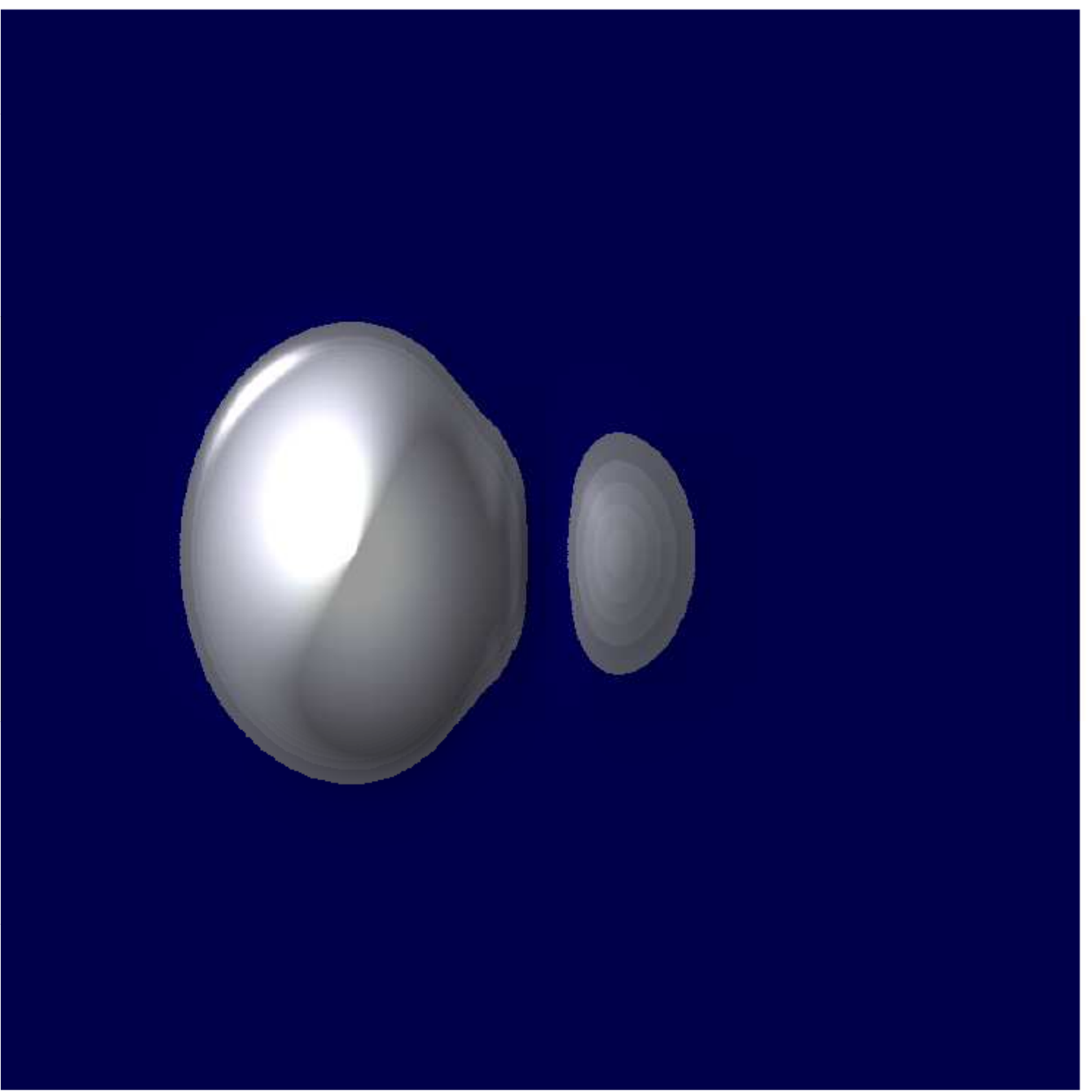}}}
{\scalebox{.35}{\includegraphics{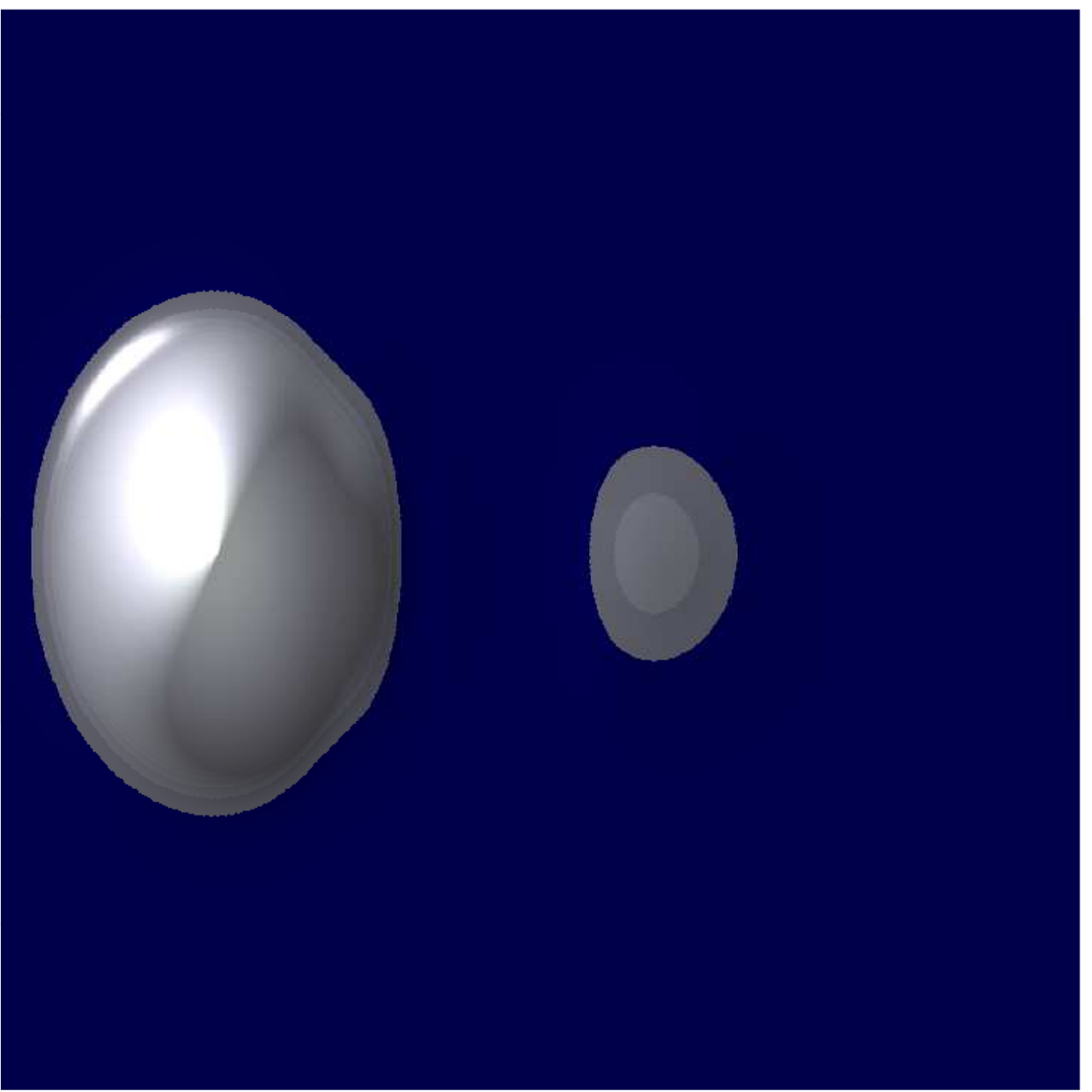}}}
{\scalebox{.35}{\includegraphics{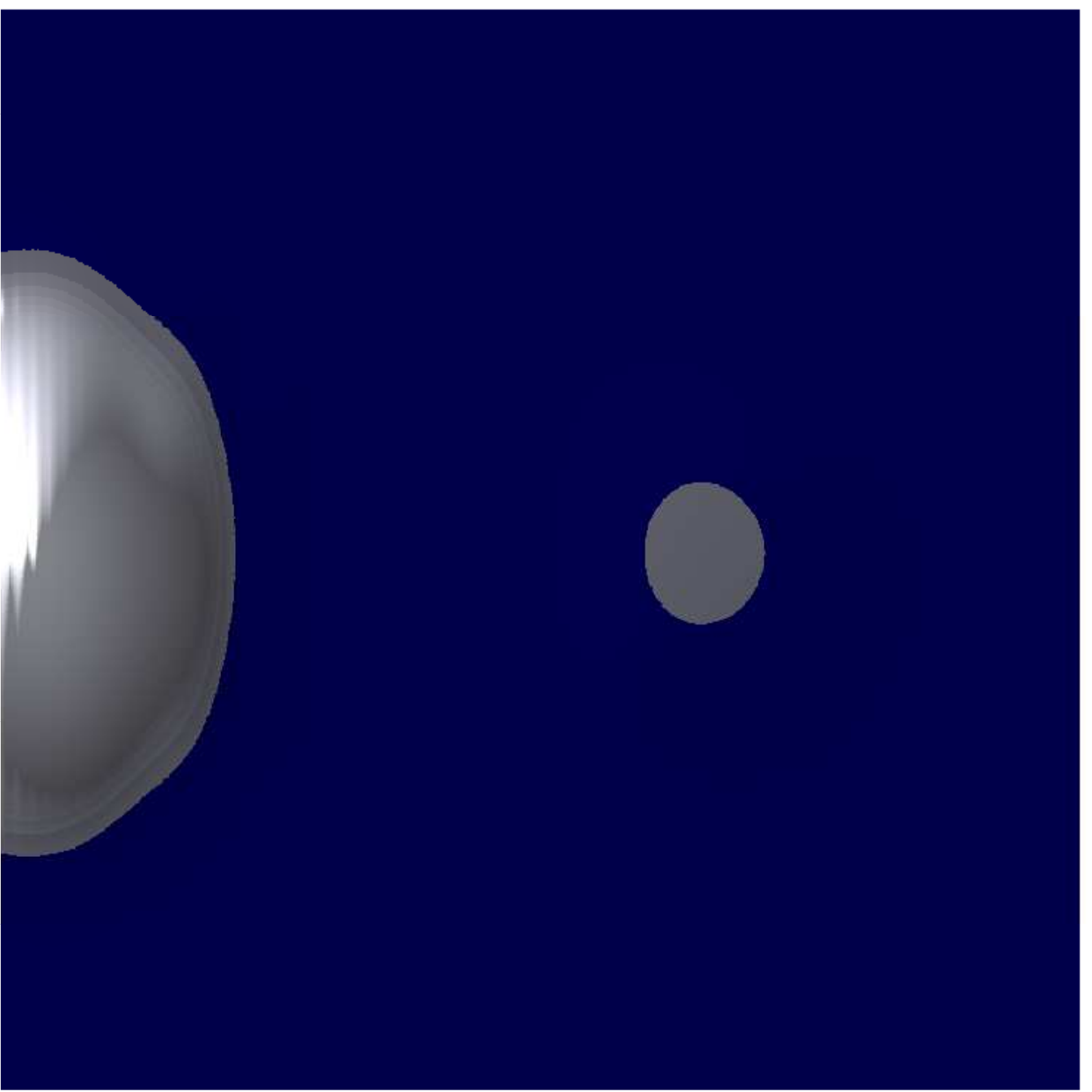}}}
\caption{\label{fig:Wave_subrepulsive} Six stages of motion of
the wave packet initialized by Eq. (\ref{Wave_packet}) in the x-z plane for the case of the 
repulsive subcritical potential step barrier. After the interference stage, the major portion of the
wave packet is reflected while a small part penetrates the barrier and dissipates. 
The animation can be accessed on-line \cite{Simi08}.}
\end{figure}

\begin{figure}
\centering
{\scalebox{.35}{\includegraphics{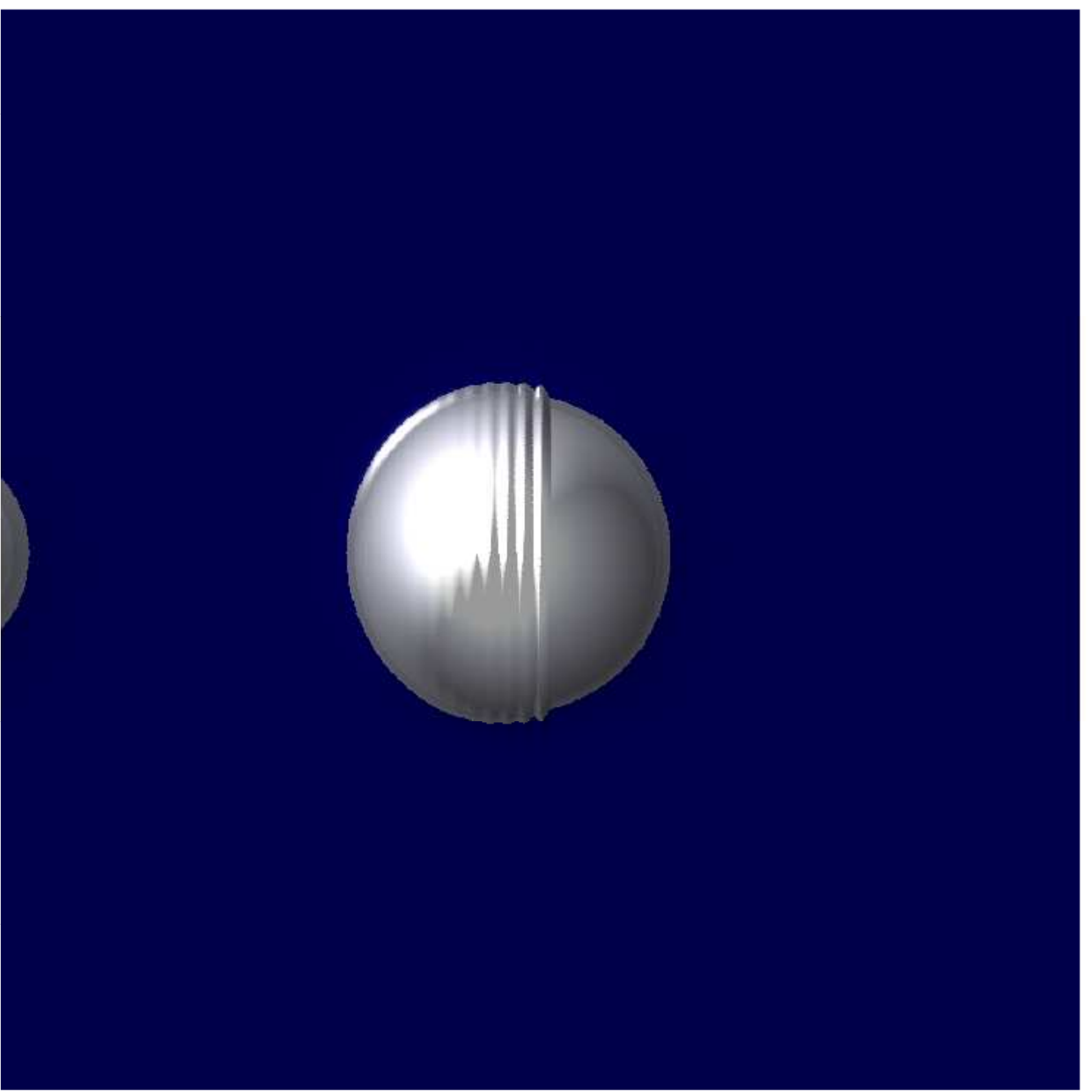}}}
{\scalebox{.35}{\includegraphics{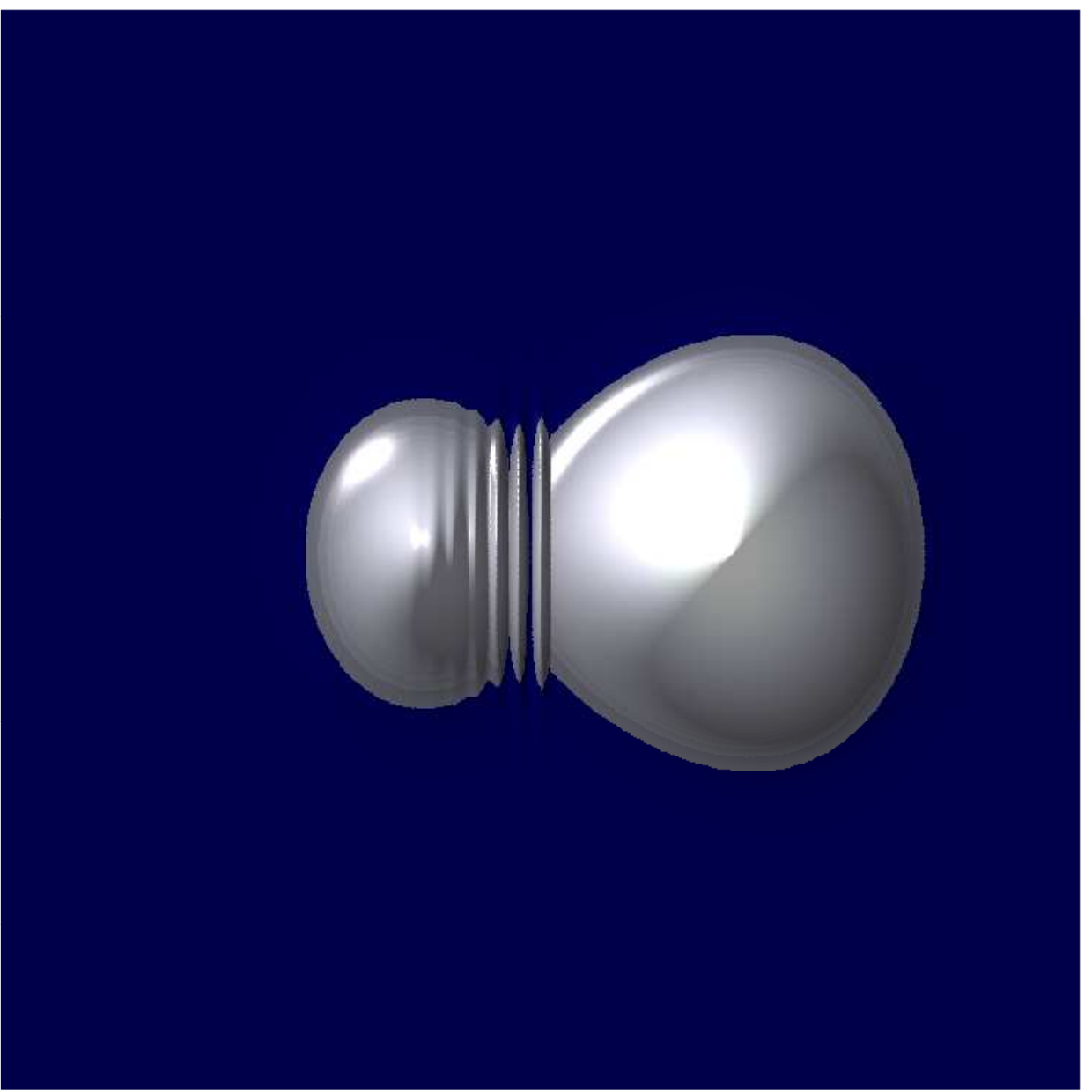}}}
{\scalebox{.35}{\includegraphics{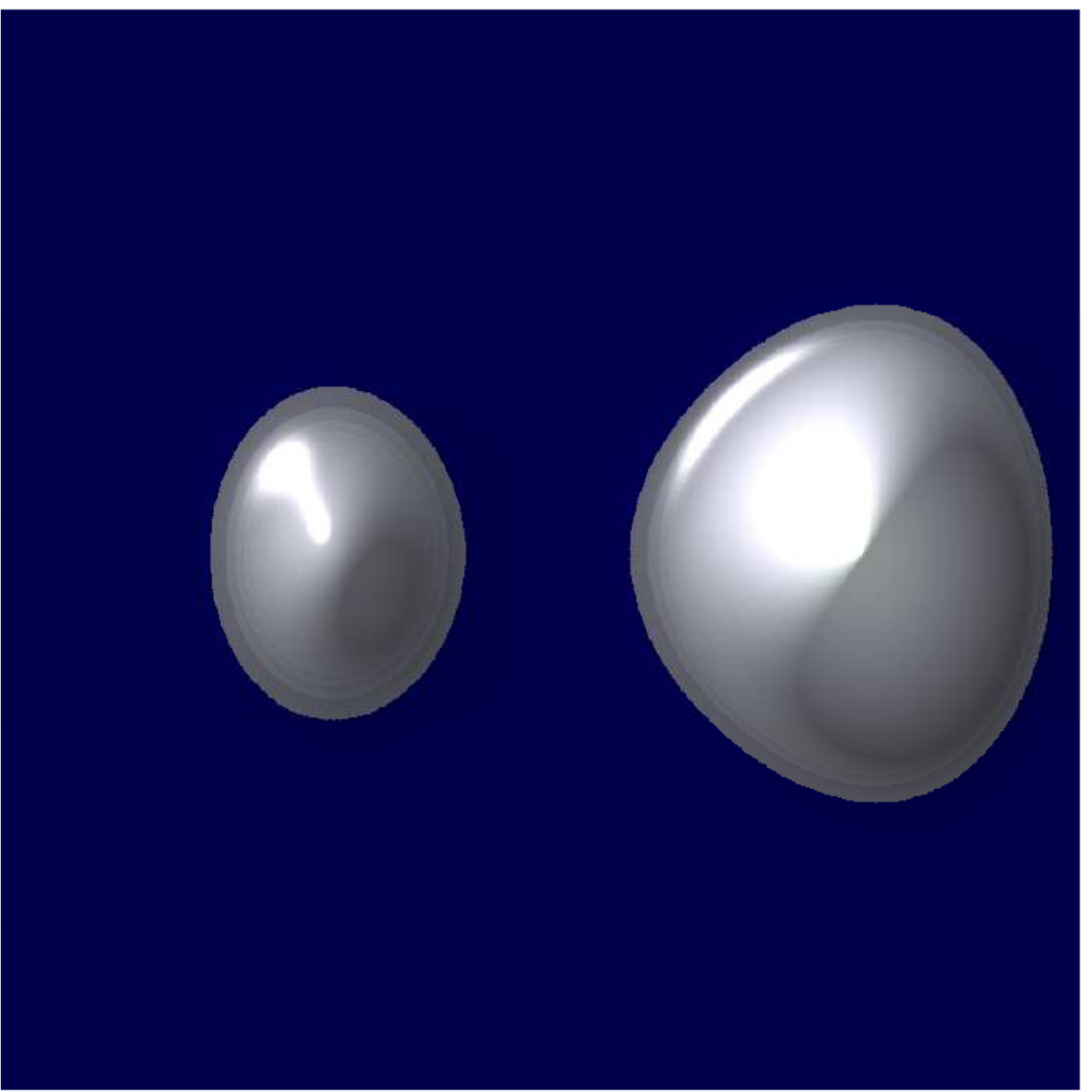}}}
\caption{\label{fig:Wave_subattractive} Three stages of motion of
the wave packet initialized by Eq. (\ref{Wave_packet}) in the x-z plane for the case of the 
attractive subcritical potential step barrier. This time, after the interference stage, 
a major portion penetrates the barrier. 
The animation can be accessed on-line \cite{Simi08}.}
\end{figure}

\begin{figure}
\centering
{\scalebox{.335}{\includegraphics{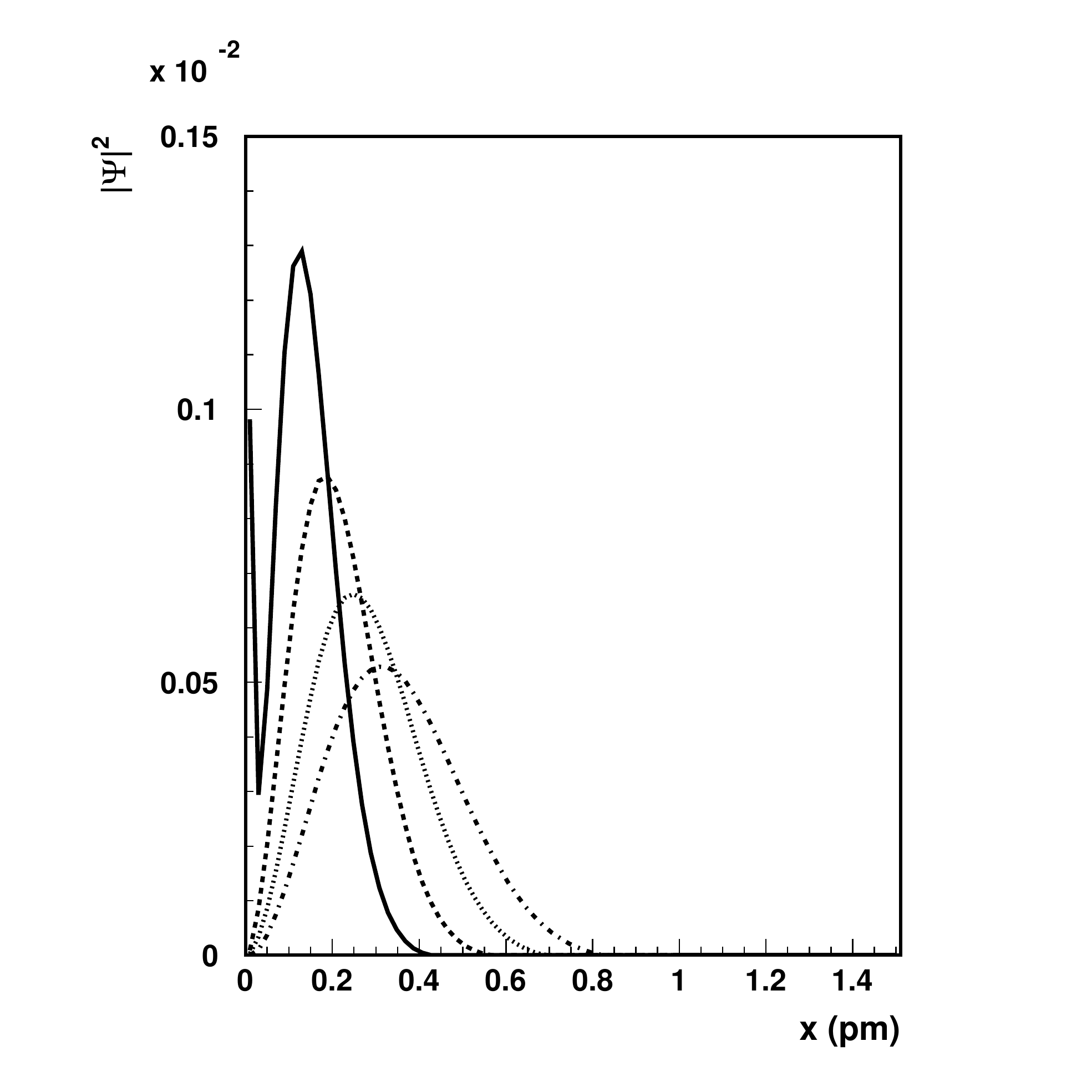}}}
{\scalebox{.335}{\includegraphics{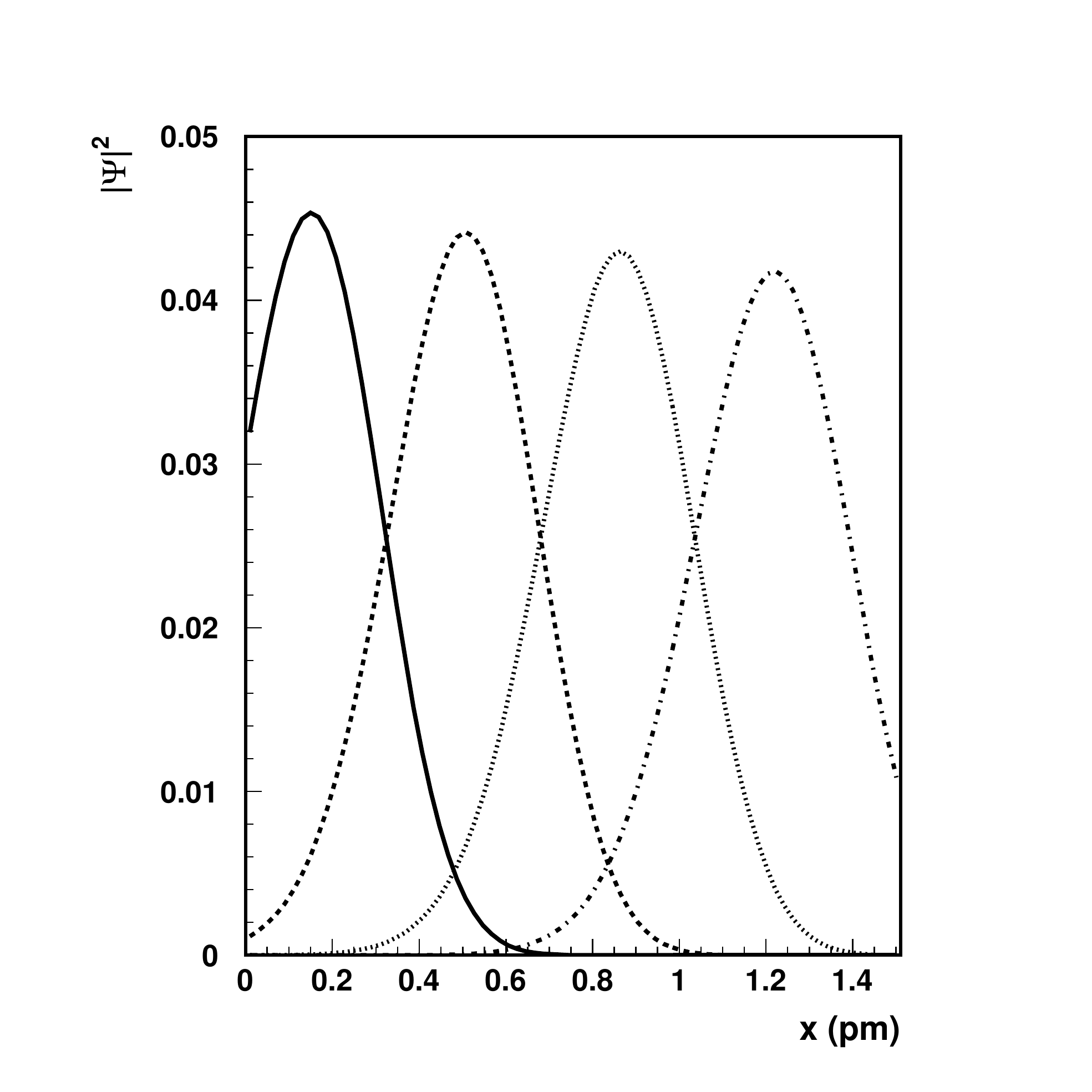}}}
\caption{\label{fig:Psi_sub_rep_att} Four time steps of the probability density function $|\Psi|^2$ 
for the dynamics 
associated with the repulsive (left)
and the attractive (right) subcritical potential step barrier. In the case of the 
repulsive potential, $|\Psi|^2$ broadens and dissipates as it penetrates the potential barrier.
In the case of the attractive potential, the distortion of $|\Psi|^2$ is small, comparable to
the distortion of a free electron wave packet \cite{Sim08}.}
\end{figure}

The first part of this paper describes the dynamics of scattering of  the wave packet initiated by 
Eq. (\ref{Wave_packet}) from the potential barrier
\begin{equation}
{A_{0} =\left\{ \begin{array} {c} V \;\;\; \mbox{ for $x \geq 0$} \\ 0 \;\;\;\;\; \mbox{ for $x < 0$}
 \end{array} \right. },
\label{Pot_barrier}
\end{equation}
where $x$  represents the coordinate in the x-direction. The initial values of the 
wave packet momenta were $p_{1} = 18.75 MeV/c$  and $p_{2}=p_{3}=0$, and the 
size was defined by $x_{0}=10^{-13} \; m$.
The step barrier repulsive potential was $V=25 \times 10^{6}\; V$,  satisfying  $eV>E+mc^{2}$.
The dynamics of the scattering of this wave packet from the repulsive potential 
is shown in Fig. \ref{fig:Wave_repulsive}.
After splitting in two as a result of the initial conditions \cite{Sim08,Huang52,Thal04},
the larger component of the wave packet moves in 
the direction of the repulsive supercritical potential step barrier and completely reflects from it. 
While we can observe the interference between the incoming and reflected waves in front of the
potential barrier, no Klein paradox is observed. The wave packet did not penetrate 
the potential barrier as predicted by the Klein solution of the Dirac equation, but completely 
reflected back as in the case of non-relativistic quantum mechanics. 
For comparison, the dynamics
associated with the attractive potential step barrier of the same magnitude is shown in 
Fig. \ref{fig:Wave_attractive}. The wave packet partially reflects and 
partially penetrates the potential step barrier. 
Integration of the probability density function $|\Psi|^2$, shown in Fig. \ref{fig:Psi_rep_att}, 
gives the reflection and transmission coefficients $R$ and $T$. In the case of  
the repulsive supercritical potential step barrier, as already discused, $R=1$ and $T=0$. 
(Increasing the step barrier repulsive potential did not change this result.) 
In the case of  the attractive supercritical potential step barrier $R=0.46$ and $T=0.54$.
The cases of attractive or repulsive subcritical potential barriers, when $eV<E+mc^{2}$,
also show nonpeculiar behavior. The same wave packet was scattered from the 
step barrier potential of $V=\pm 50 \times 10^{5} \; V$. The dynamics of the scattering 
from the repulsive potential is shown in Fig. \ref{fig:Wave_subrepulsive} and 
from the attractive potential in Fig. \ref{fig:Wave_subattractive}. 
In the case of the repulsive potential after the interference at the potential barrier, part of the
wave packet is reflected and part penetrates the barrier. The wave packet which 
has penetrated the potential barrier broadened and dissipated, showing, as expected 
in the non-relativistic case, damping with distance. In the case of the 
attractive potential part of the wave packet which has penetrated the potential 
barrier showed little broadening, comparable to the distortion of the free electron 
wave packet \cite{Sim08}. The shapes of the wave packets after the penetration 
into the potential barriers are shown in  Fig. \ref{fig:Psi_sub_rep_att}.
\begin{figure}
\centering
{\scalebox{.35}{\includegraphics{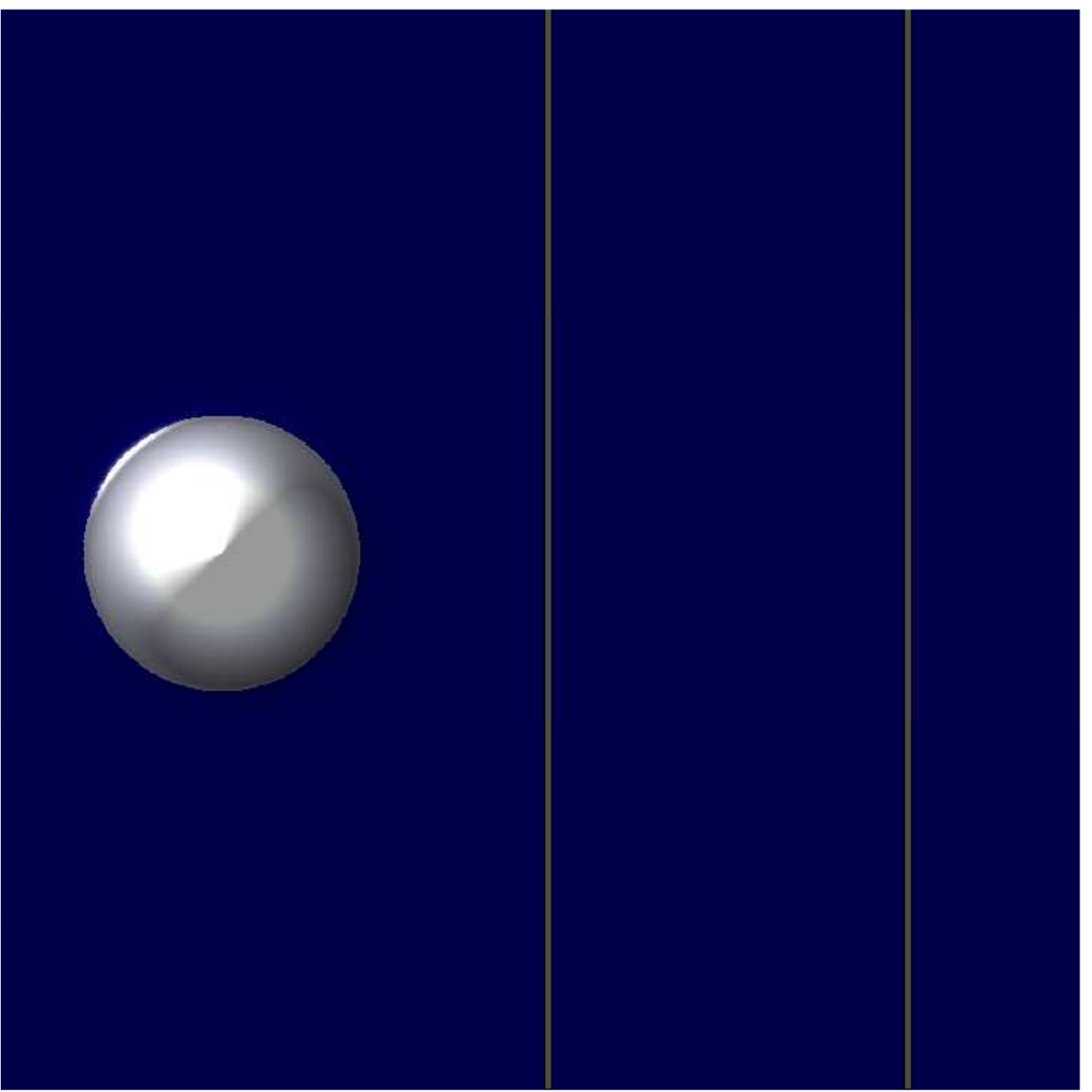}}}
{\scalebox{.35}{\includegraphics{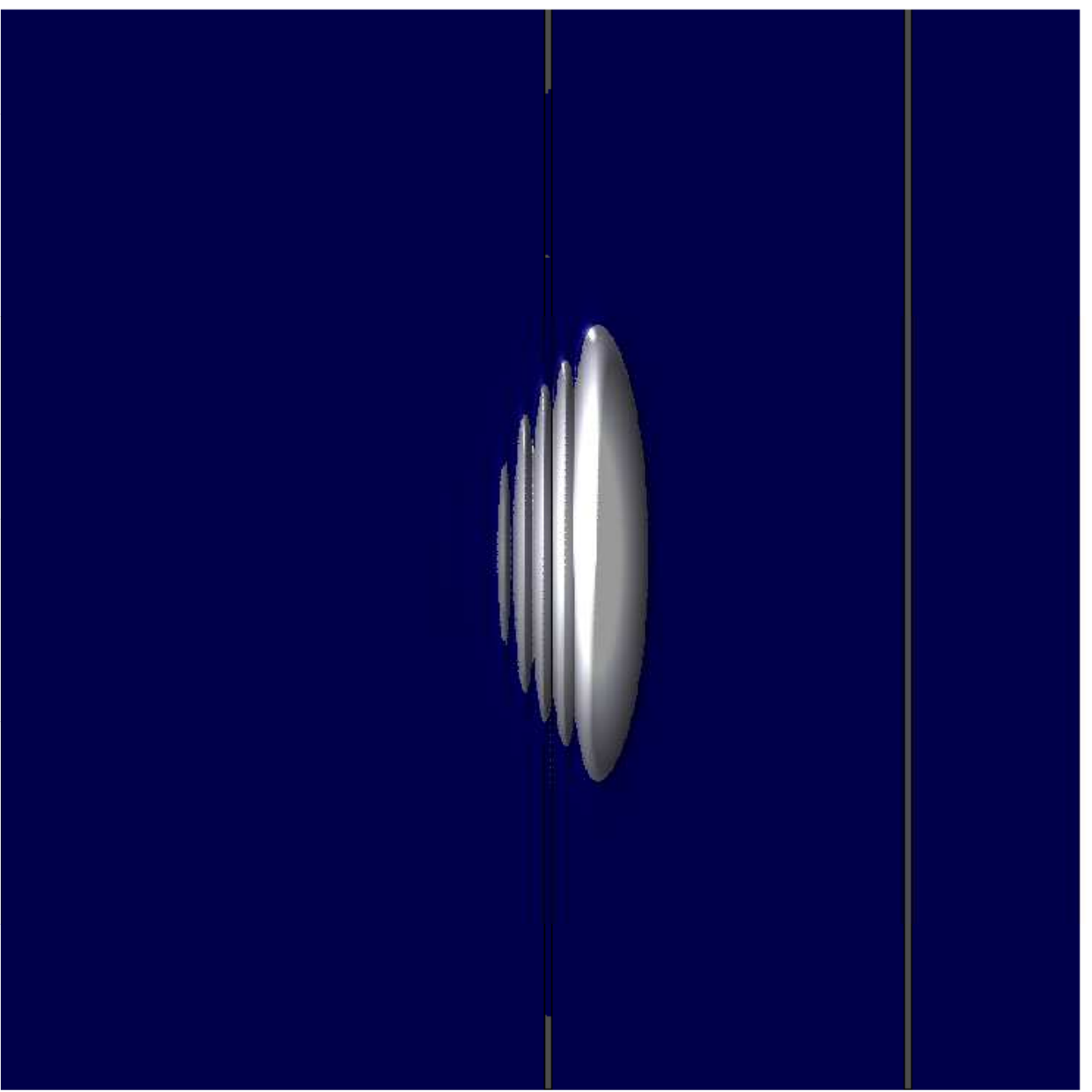}}}
{\scalebox{.35}{\includegraphics{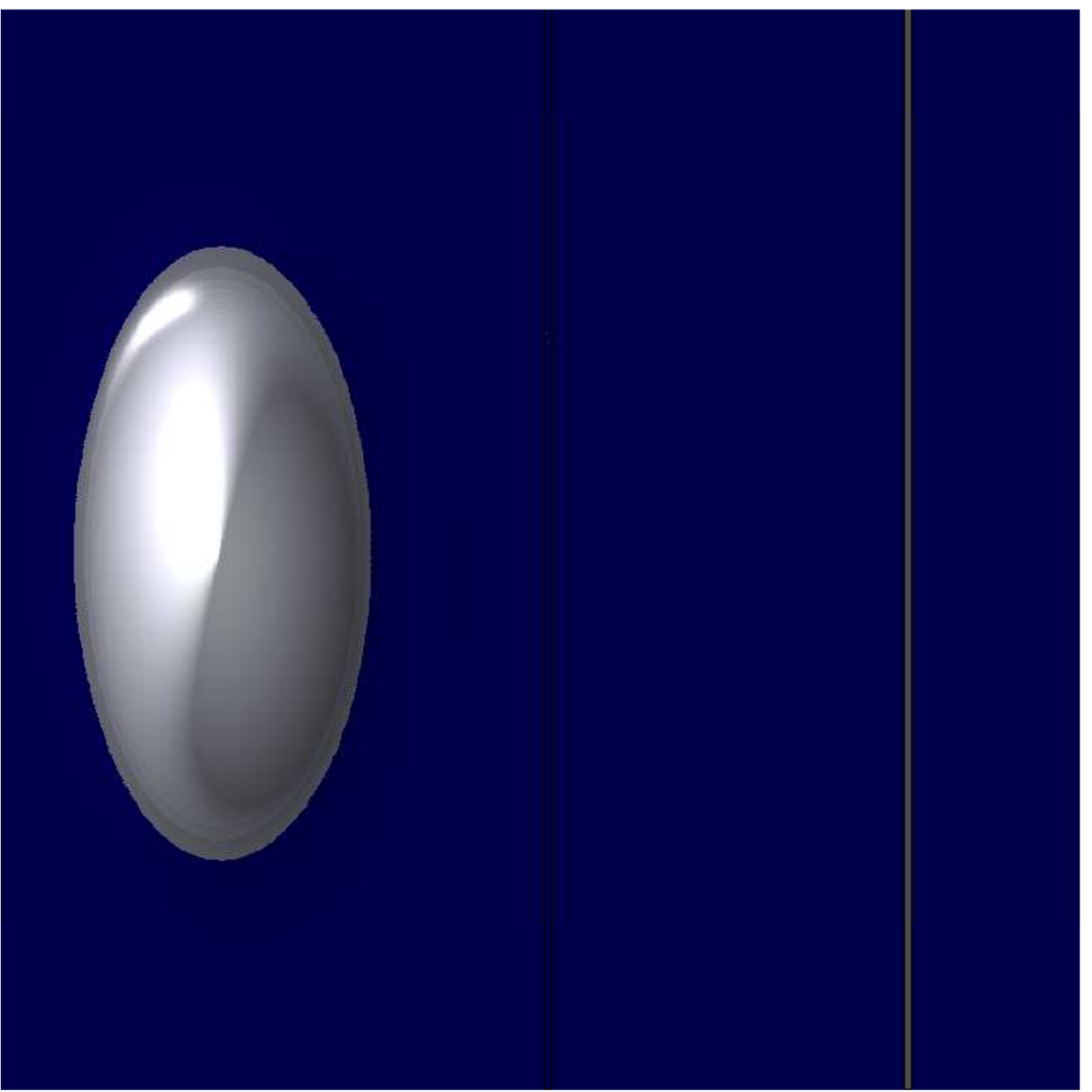}}}
\caption{\label{fig:Wave_Const_field} Three stages of motion of
the wave packet initialized by Eq. (\ref{Wave_packet}) in the x-z plane for the case of the 
repulsive potential barrier described by Eq. (\ref{Pot_barrier2}). 
Left and right figures correspond to the wave packet before and after the scattering. 
Middle figure corresponds to the maximal penetration. 
The lines correspond to the region $a \geq x \geq 0$ in Eq. (\ref{Pot_barrier2}).
The animation can be accessed on-line \cite{Simi08}.}
\end{figure}
\begin{figure}
\centering
{\scalebox{.35}{\includegraphics{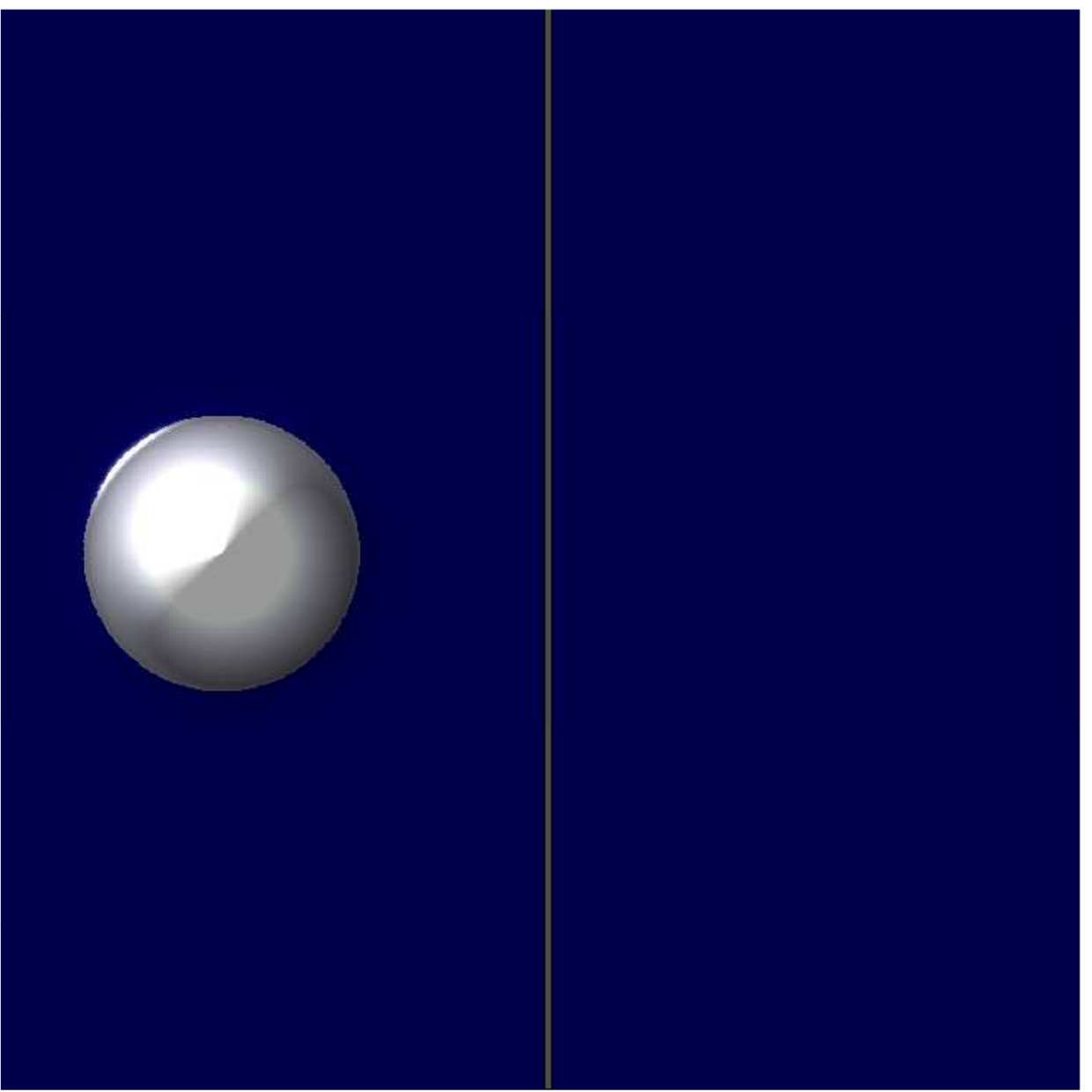}}}
{\scalebox{.35}{\includegraphics{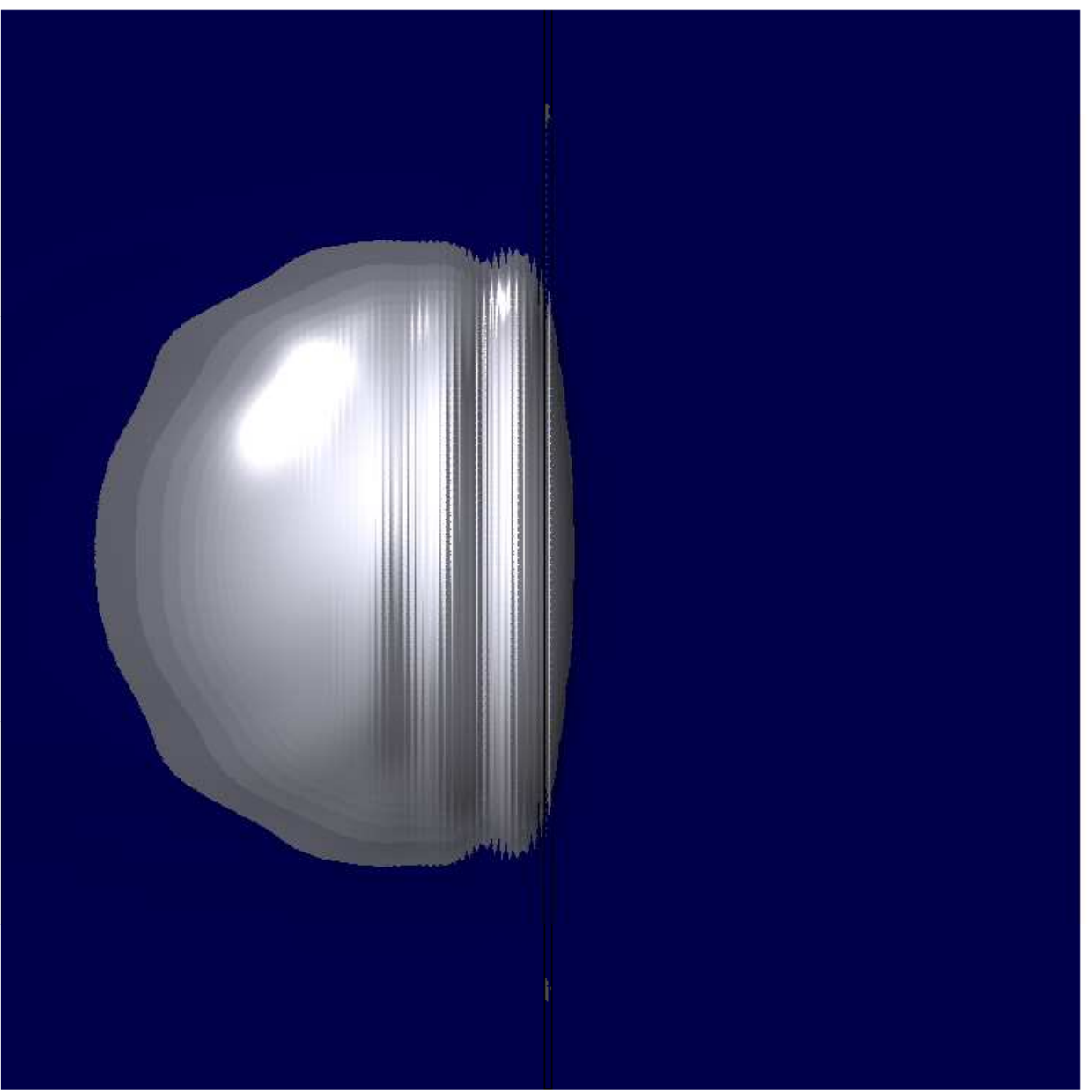}}}
{\scalebox{.35}{\includegraphics{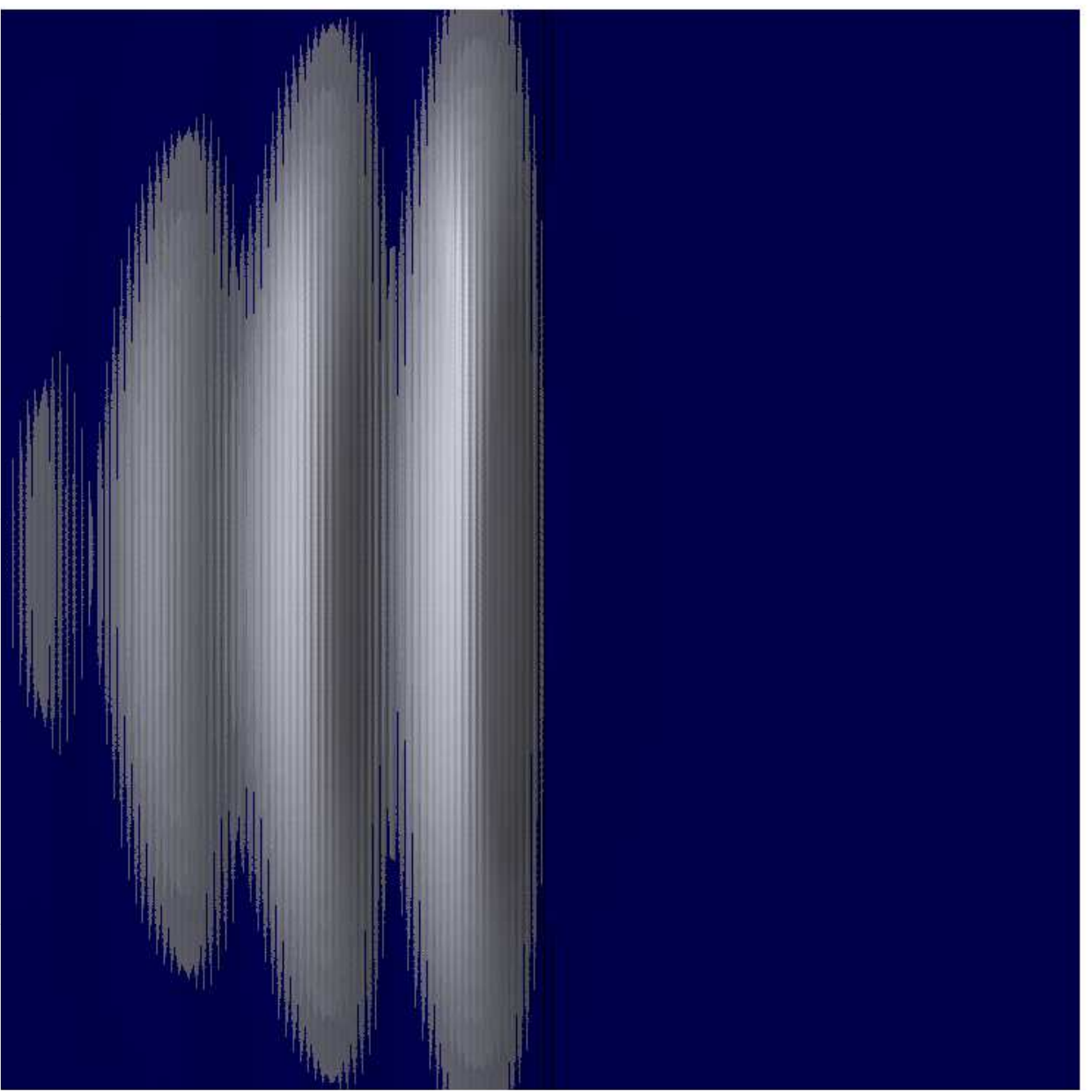}}}
{\scalebox{.35}{\includegraphics{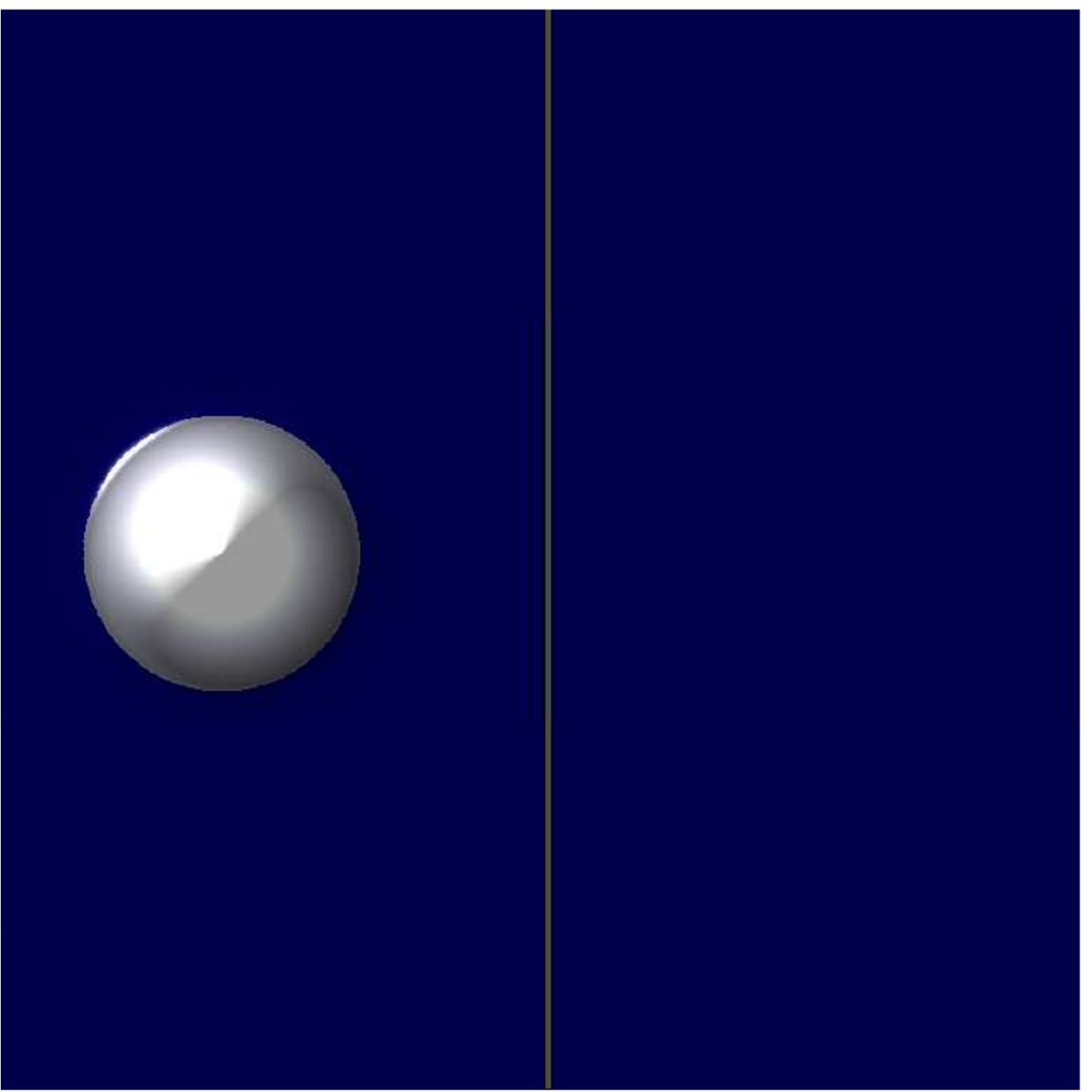}}}
{\scalebox{.35}{\includegraphics{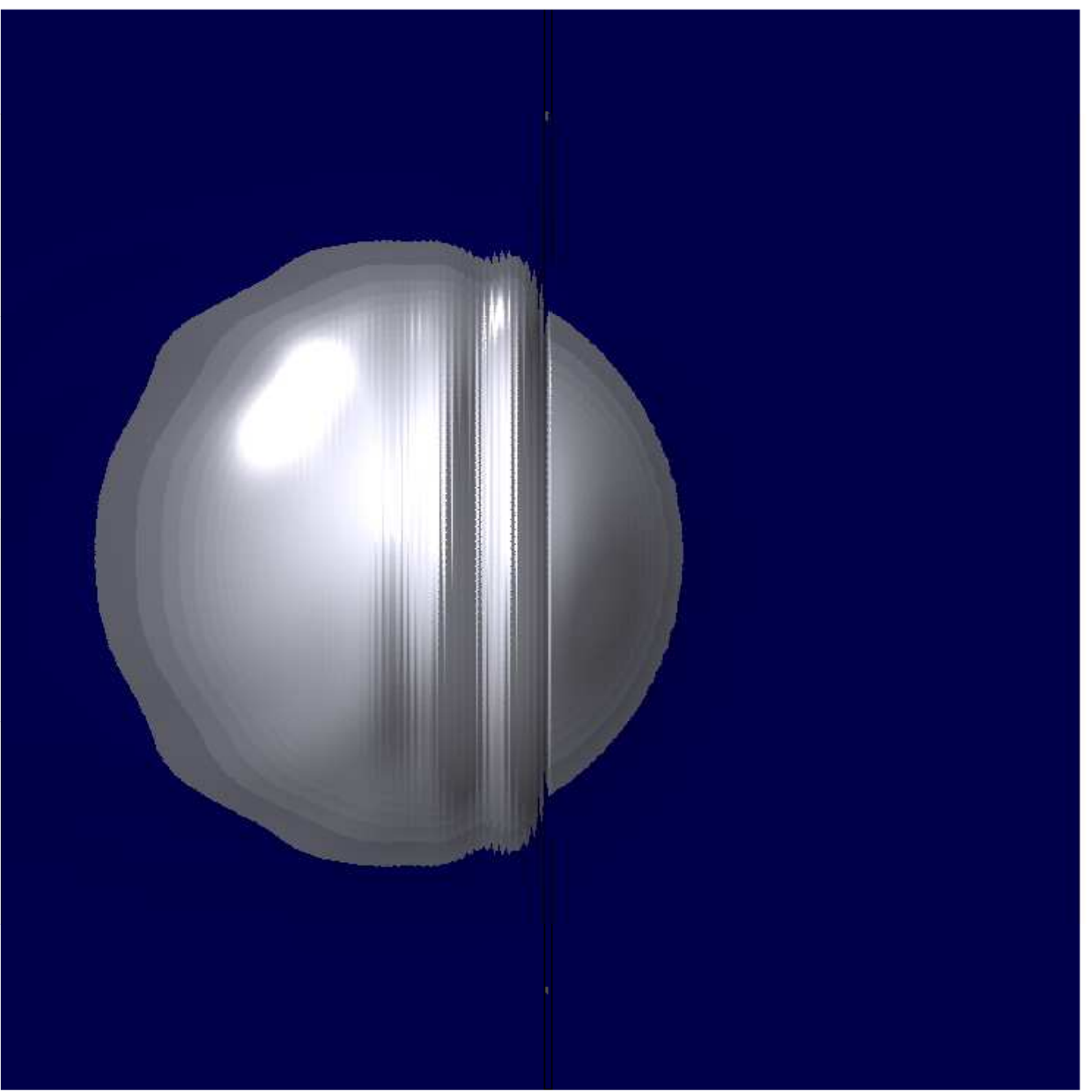}}}
{\scalebox{.35}{\includegraphics{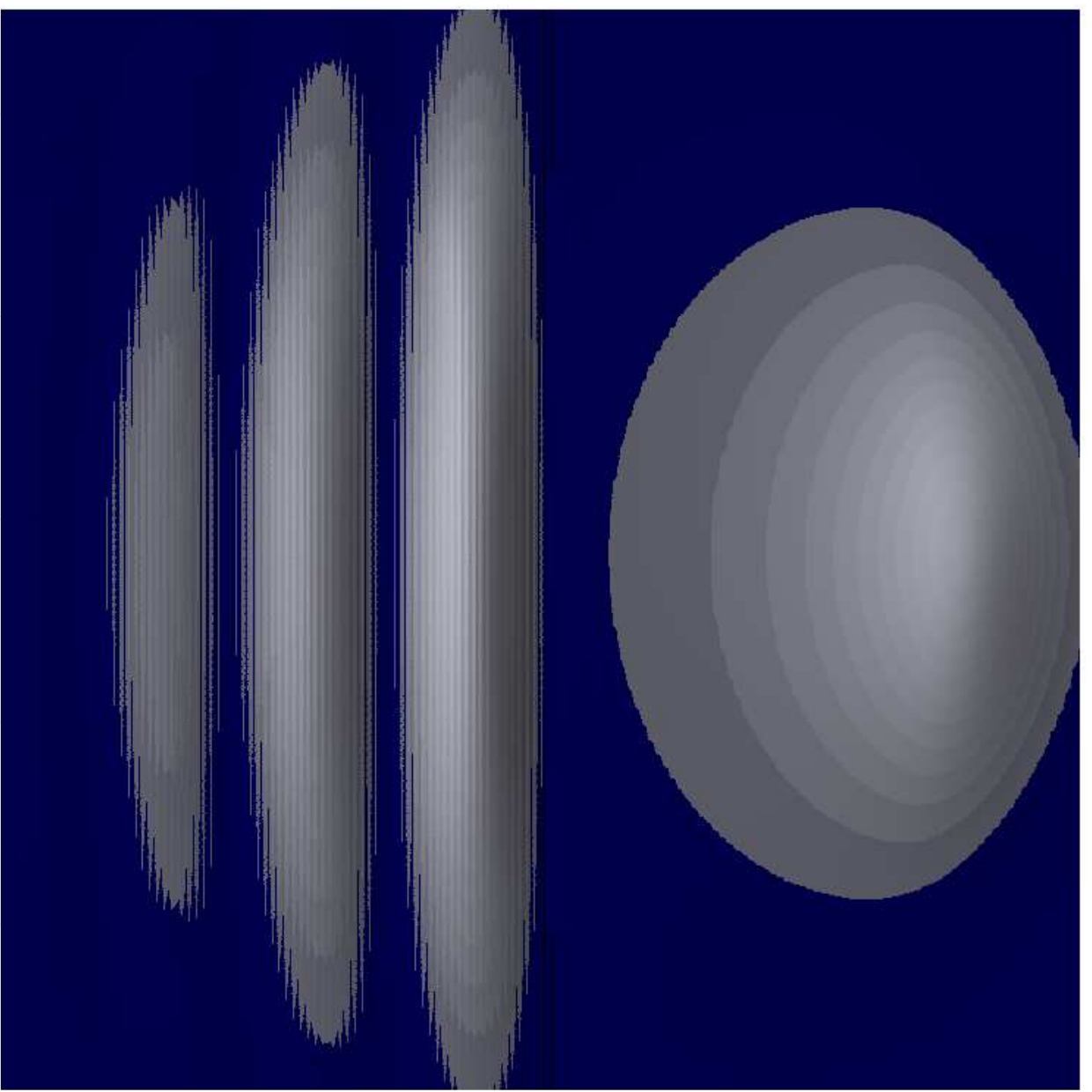}}}
\caption{\label{fig:Wave_low_momentum} Three stages of motion, in the x-z plane, of
the wave packet of the momentum $p_{1} = 0.1875 MeV/c$  and $p_{2}=p_{3}=0$
initialized by Eq. (\ref{Wave_packet}). Top row is for the supercritical and 
bottom row  for the subcritical repulsive potential. Left and right figures correspond to the 
wave packet before and after the scattering. Middle figure corresponds to 
the time of interaction of the wave packet with the potential barrier. The line corresponds to $x = 0$.
The animation can be accessed on-line \cite{Simi08}.}
\end{figure}
To test the effect of the shape of the potential 
barrier and the effect of the wave packet shape and momentum, two more tests were performed.
To verify that the properties of the wave packet propagation are not due to the 
shape of the potential barrier, the step potential barrier,  Eq. (\ref{Pot_barrier}), was replaced 
by the potential representing a constant electric field in a finite region of space 
\begin{equation}
{A_{0} =\left\{ \begin{array} {c} V \;\;\; \mbox{ for $x > a$}
\\ -\varepsilon x \;\;\;\;\; \mbox{ for $a \geq x \geq 0$} 
\\ 0 \;\;\;\; \mbox{ for $x < 0$}
 \end{array} \right. }.
\label{Pot_barrier2}
\end{equation}
In this particular case, $a =50 \times 10^{-14}\; m$, $\varepsilon =-50 \times 10^{18}\; V/m$,
and $V=25 \times 10^{6}\; V$.  $x$  represents the coordinate in the x-direction.
The snapshots of the dynamics of the scattering of the same wave packet, 
Eq. (\ref{Wave_packet}), from the repulsive potential of this form are shown
 in Fig. \ref{fig:Wave_Const_field}. Again, the wave packet did not penetrate the
potential and no Klein paradox was observed.
\begin{figure}
\centering
{\scalebox{.35}{\includegraphics{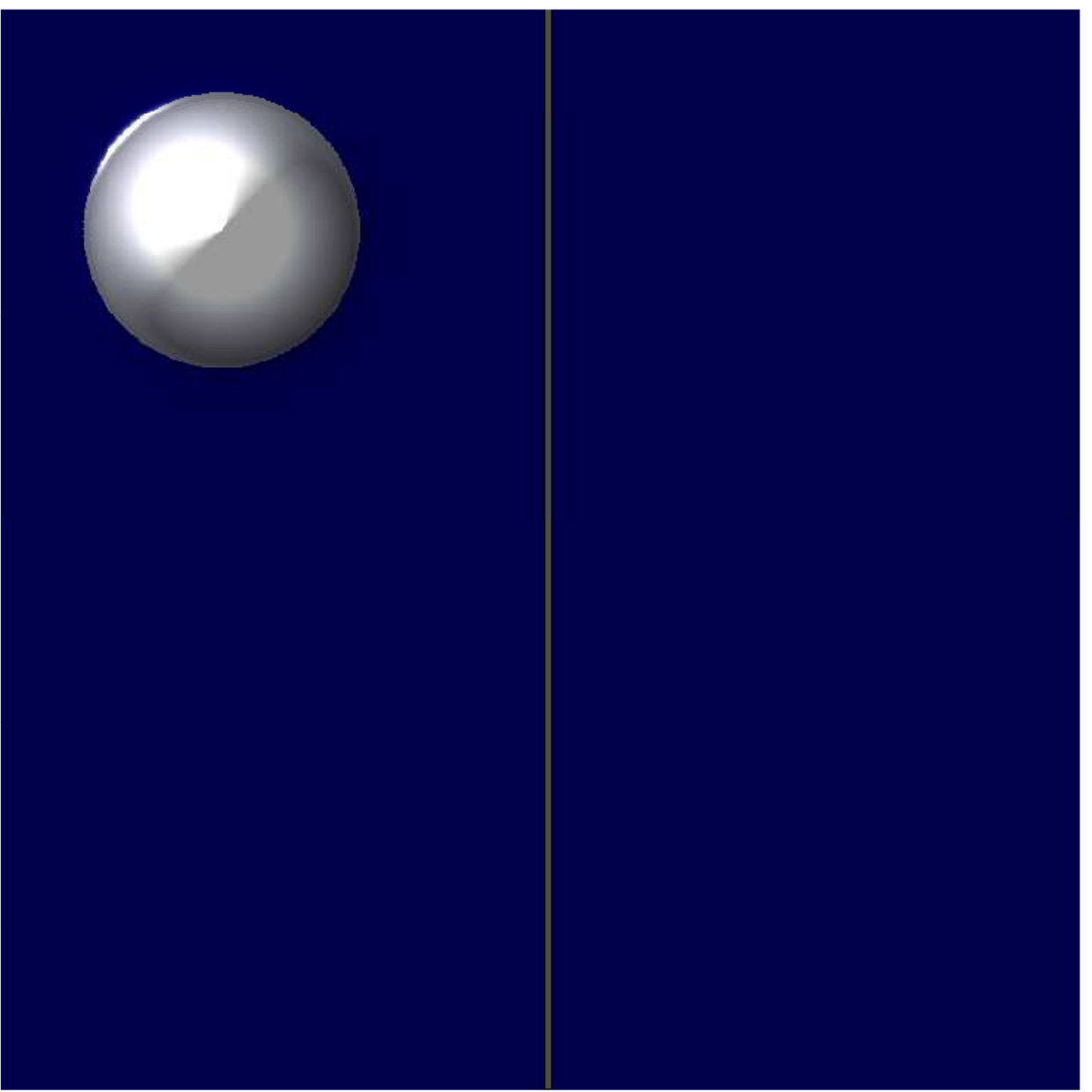}}}
{\scalebox{.35}{\includegraphics{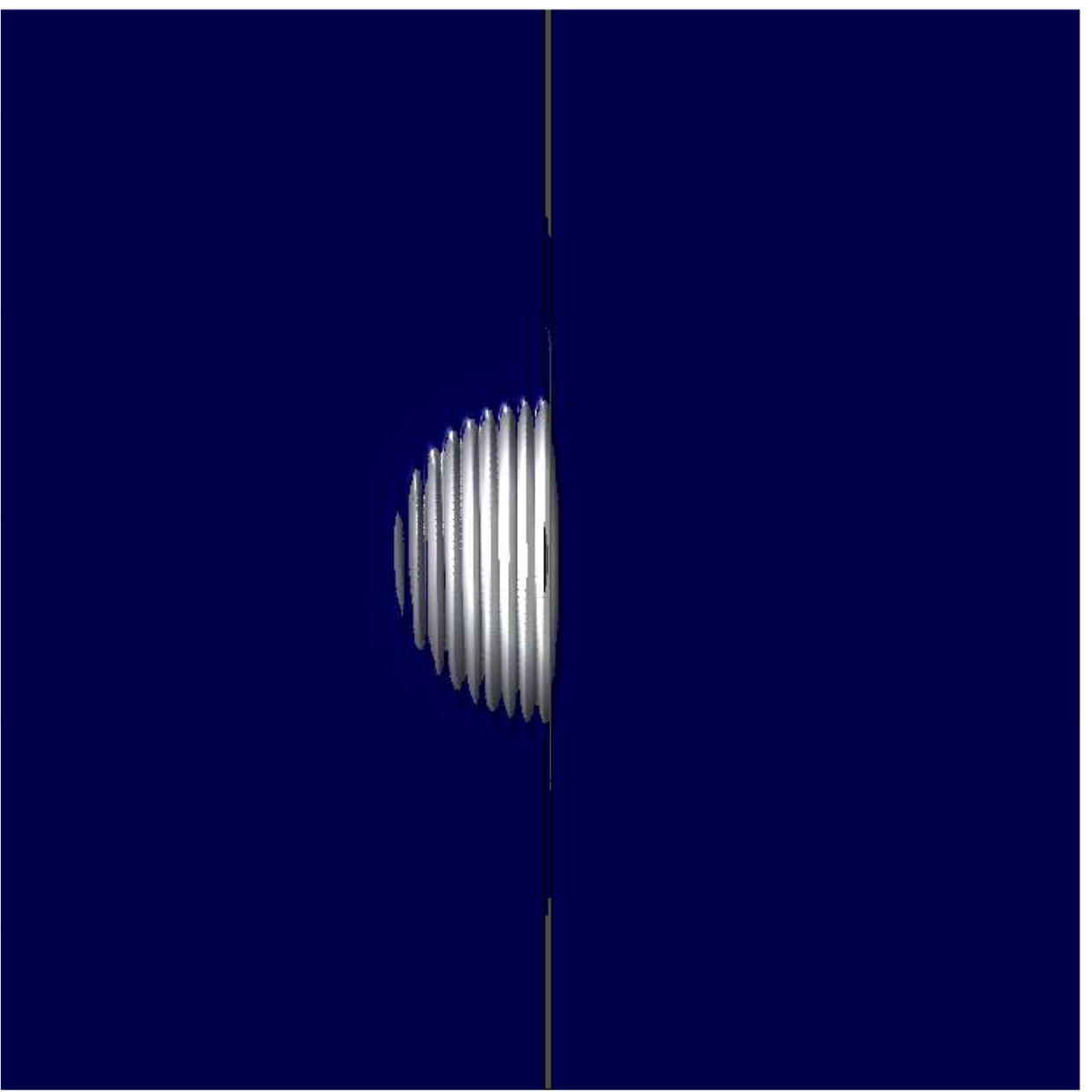}}}
{\scalebox{.35}{\includegraphics{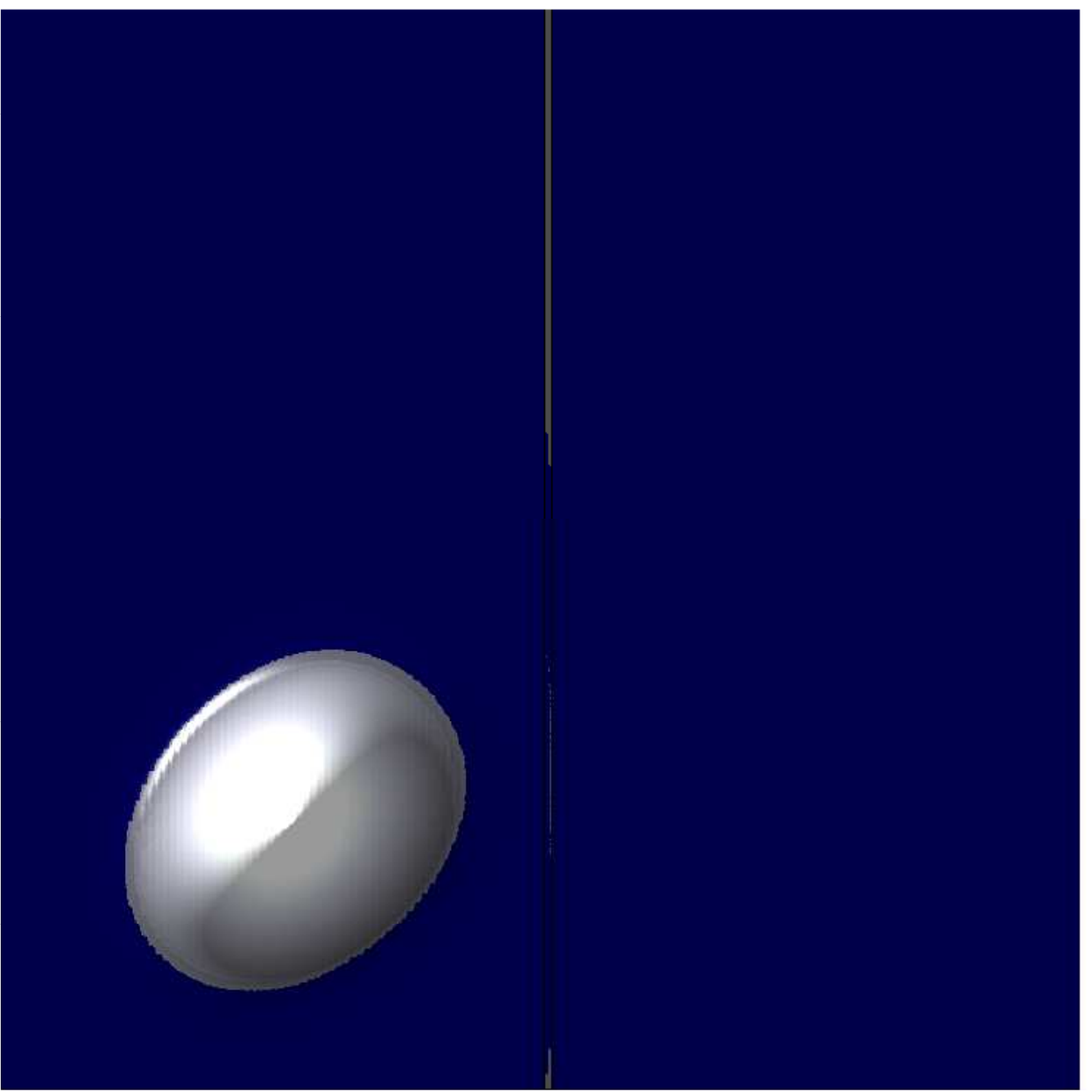}}}
{\scalebox{.35}{\includegraphics{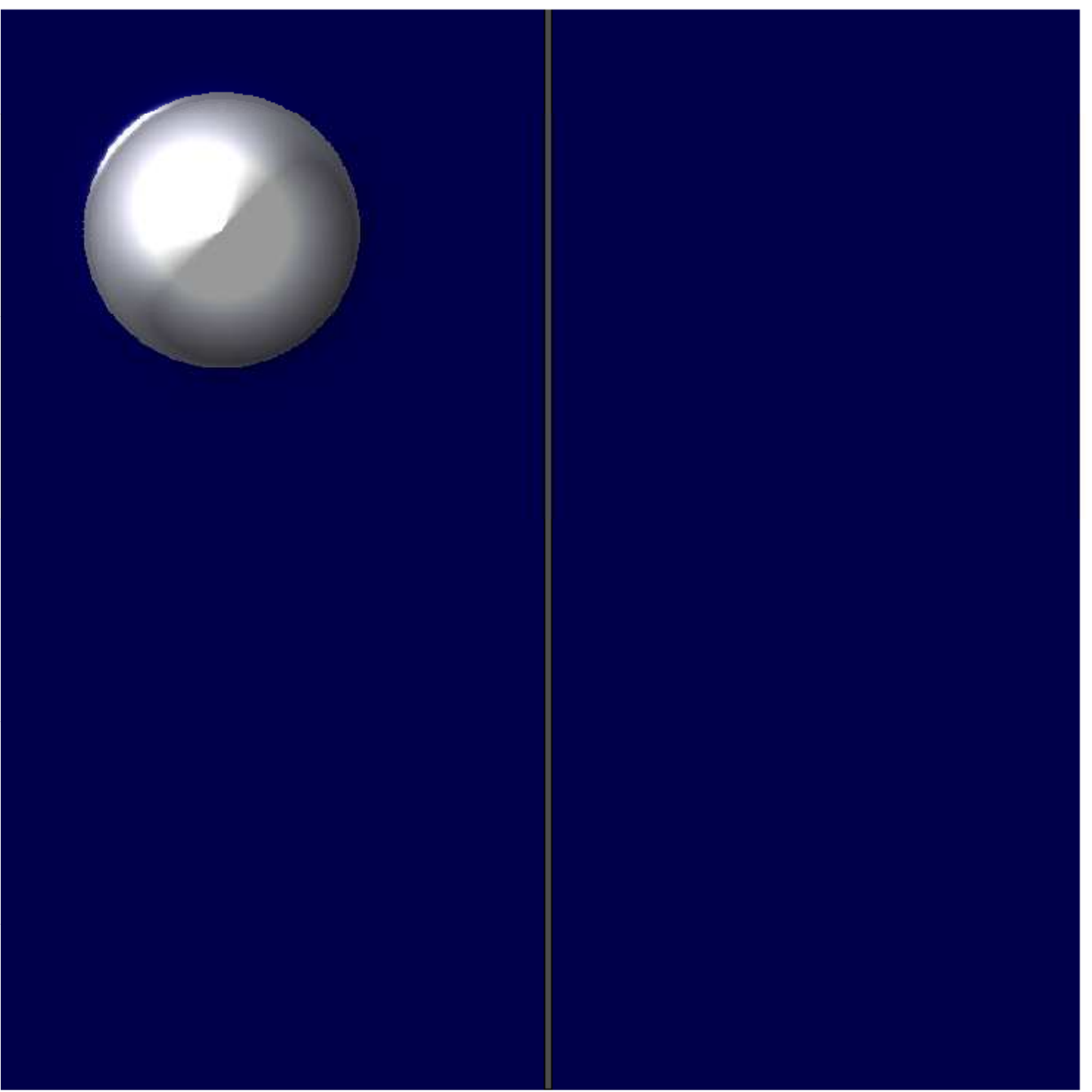}}}
{\scalebox{.35}{\includegraphics{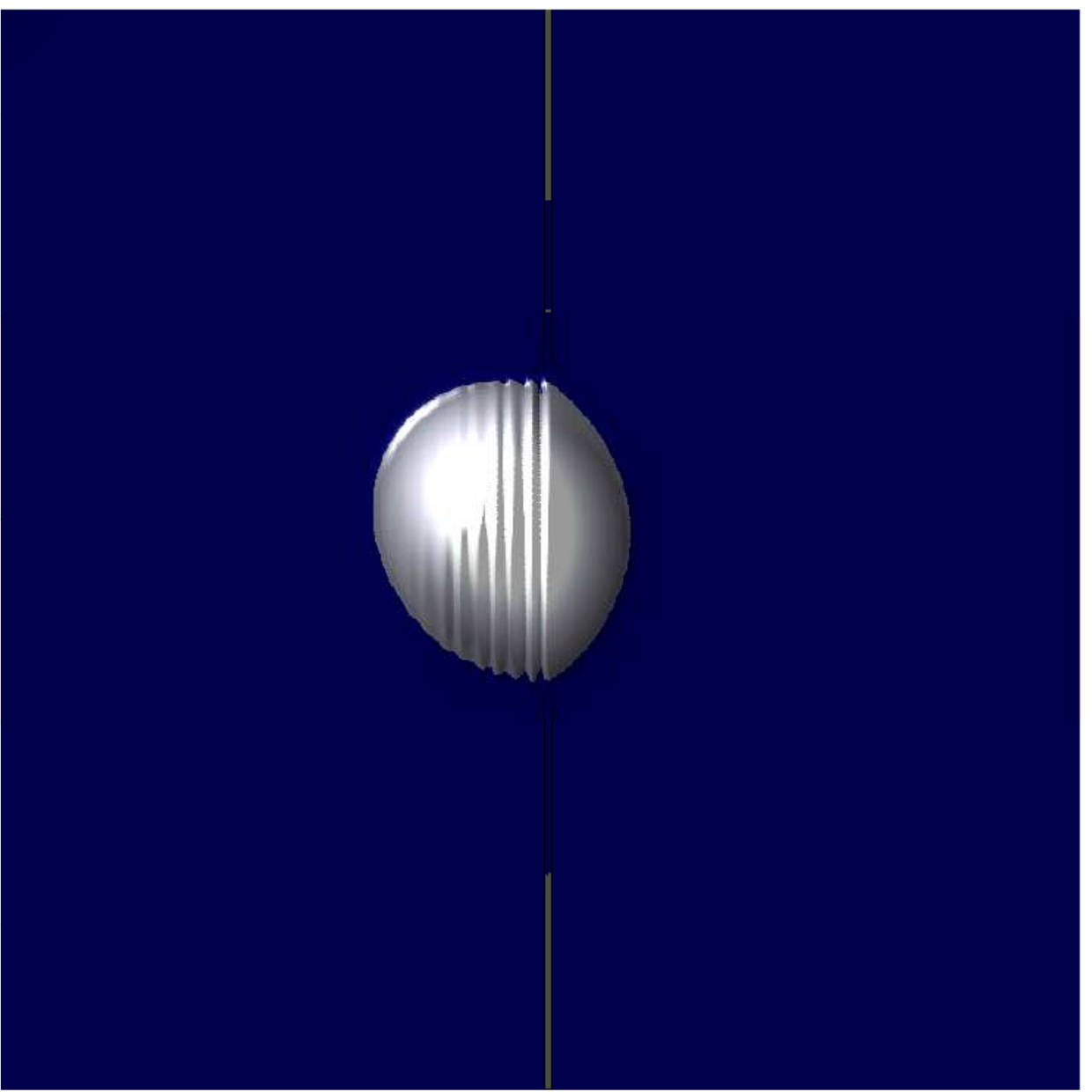}}}
{\scalebox{.35}{\includegraphics{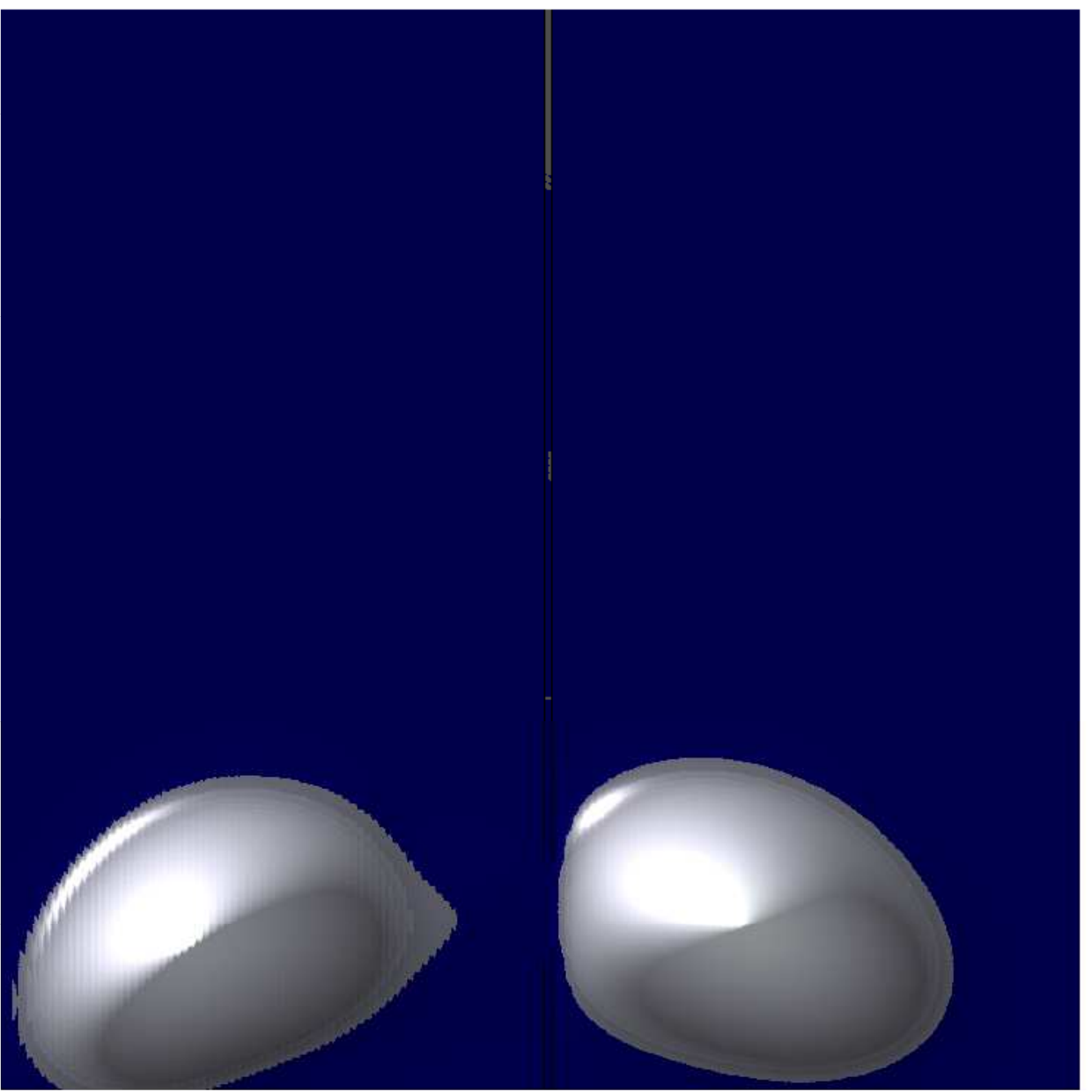}}}
\caption{\label{fig:Wave_45deg} Three stages of motion of
the wave packet in the x-z plane for the case of the angle of incidence of $45^{\circ}$.
Top row is for the repulsive and the bottom row for attractive supercritical potential step barrier.
In the case of the repulsive potential the wave packet reflects from the potential
with the reflected angle equal to the incident angle.  In the case of the attractive potential, 
part of the wave packet reflects from and part refracts into the potential. 
The animation can be accessed on-line \cite{Simi08}.}
\end{figure}

The importance of the wave packet momentum and shape was tested by reducing the initial momentum
to $p_{1} = 0.1875 MeV/c$  and $p_{2}=p_{3}=0$, and increasing the width of the wave 
packet to  $x_{0}=10^{-12} m$. This reduced the momentum due to the particle 
localization to $\Delta p \sim \hbar / \Delta x \sim 0.2 MeV/c$, keeping the energy of the
wave packet below particle production threshold and preventing generation of appreciable 
amounts of negative-energy components of the wave packet \cite{Sim08,Sak87}.
The step barrier supercritical repulsive potential, satisfying  $eV>E+mc^{2}$, 
was chosen $V=1.5 \times 10^{6}\; V$, and subcritical repulsive potential, 
satisfying  $eV<E+mc^{2}$,  was chosen $V=1.0 \times 10^{6}\; V$. The snapshots 
of the dynamics of the scattering in both cases are shown in 
Fig. \ref{fig:Wave_low_momentum}. The wave packet did not penetrate the
supercritical repulsive potential and no Klein paradox was observed. The wave packet did 
penetrate the subcritical repulsive potential with the reflection and transmission 
coefficients of $R \simeq 0.6$ and $T \simeq 0.4$

We also tested the effect of the angle of impact. The initial values of the 
wave packet momenta were $p_{1} = p_{3} = 18.75 MeV/c$  and $p_{2}=0$, and the 
step barrier repulsive potential was $V=25 \times 10^{6}\; V$. Such a wave packet
scatters from the potential at an angle of $45^{\circ}$. The dynamics of this
scattering is shown in Fig. \ref{fig:Wave_45deg}.
The wave packet reflects from the potential similarly to the reflection of the
electromagnetic wave with the reflected angle equal to the incident angle. 
The wave packet did not penetrate the potential barrier and, again, 
the Klein paradox was not observed. When the repulsive potential was replaced with
an attractive potential, part of the wave packet reflected and part refracted into the potential. 
This scattering is also shown in Fig. \ref{fig:Wave_45deg}.

Finally, to test spin orientation, some of the studies were repeated for the dynamics
of the spin-flipped wave packet defined by its initial wave function
\begin{equation}
{\Psi (\vec x,0) =N \sqrt{\frac{E+mc^{2}}{2E}}\left( \begin{array} {c} 0 \\ 1
\\ \frac{(p_{1}-ip_{2})c}{E+mc^{2}} \\ \frac{-p_{3}c}{E+mc^{2}}\end{array} \right)}
e^{-\frac{\vec x \cdot \vec x }{4x_{0}^{2}}+\frac{i\vec p \cdot \vec x}{\hbar}}.
\label{Wave_packet2}
\end{equation}
The results did not change.

To conclude, the full three-dimensional Finite Difference Time Domain (FDTD)
method was developed to solve the Dirac equation. In this paper, the method was
applied to the dynamics of a Dirac electron in a scalar potential, particularly to the
arrangements corresponding to the dynamics associated with the Klein paradox.
The Klein paradox, resulting from the solution of the stationary Dirac equation,  predicts 
unimpeded penetration of particles in the energy forbidden region, in contradiction 
with intuition and the behavior described by Schrodinger equation. 
Solutions of the time-dependent Dirac equation presented in this paper 
show no such penetration leading
to the conclusion that the Klein paradox does not exist.

I would like to thank Dentcho Genov, B. Ramu Ramachandran, Lee Sawyer, Ray Sterling and Steve Wells 
for useful comments.
Also, the use of the high-performance computing resources provided by Louisiana Optical
Network Initiative (LONI; www.loni.org) is gratefully acknowledged.

\end{document}